\documentclass[12pt]{article}
\usepackage[letterpaper,margin=1in]{geometry}
\usepackage{setspace}
\usepackage{hyperref}
\usepackage{natbib}
\usepackage{amsmath,mathrsfs,amsthm,graphicx,bm,amssymb}
\usepackage{algorithm}
\usepackage{color}
\usepackage{ulem}                           
\usepackage{enumitem}
\usepackage{multirow}
\usepackage{pdflscape}
\newtheorem{theorem}{Theorem}[section]
\newtheorem{lemma}{Lemma}[section]
\newtheorem{definition}{Definition}[section]
\newtheorem{corollary}{Corollary}[section]
\DeclareMathOperator*{\argmin}{argmin}

\newcommand{\blind}{1}
\begin{document}

\def\spacingset#1{\renewcommand{\baselinestretch}%
{#1}\small\normalsize} \spacingset{1}

\setlength{\belowdisplayskip}{0pt} \setlength{\belowdisplayshortskip}{0pt}
\setlength{\abovedisplayskip}{0pt} \setlength{\abovedisplayshortskip}{0pt}


\if1\blind
{
  \title{\bf Multi-task Learning with  High-Dimensional Noisy Images}
 \author{Xin Ma \footnote{Department of Biostatistics and Bioinfomatics, Emory University}, Suprateek Kundu \footnote{Department of Biostatistics, The University of Texas at MD Anderson Cancer Center} \footnote{Corresponding author: Email: SKundu2@mdanderson.org; Address: 1400 Pressler Street, Unit 1411, Houston, TX 77030} \\
 and\\
for the Alzheimer’s Disease Neuroimaging Initiative 
}
  \date{}
  \maketitle
} \fi

\if0\blind
{
  \bigskip
  \bigskip
  \bigskip
  \begin{center}
    {\LARGE\bf Multi-task Learning with  High-Dimensional Noisy Images}
\end{center}
  \medskip
} \fi

\begin{abstract}
Recent medical imaging studies have given rise to distinct but inter-related datasets corresponding to multiple experimental tasks or longitudinal visits. Standard scalar-on-image regression models that fit each dataset separately are not equipped to leverage information across inter-related images, and existing multi-task learning approaches are compromised by the inability to account for the noise that is often observed in images. We propose a novel joint scalar-on-image regression framework involving wavelet-based image representations with grouped penalties that are designed to pool information across inter-related images for joint learning, and which explicitly accounts for noise in high-dimensional images via a projection-based approach. In the presence of non-convexity arising due to noisy images, we derive non-asymptotic error bounds under non-convex as well as convex grouped penalties, even when the number of voxels increases exponentially with sample size. A projected gradient descent algorithm is used for computation, which is shown to approximate the optimal solution via well-defined non-asymptotic optimization error bounds under noisy images. Extensive simulations and application to a motivating longitudinal Alzheimer's disease study illustrate significantly improved predictive ability and greater power to detect true signals, that are simply missed by existing methods without noise correction due to the {\it attenuation to null} phenomenon.
\end{abstract}
\noindent%
\vfill

\newpage
\spacingset{1.5} 

\section{Introduction}
\label{sec:intro}
Methods for functional data analysis \citep{ramsay1991some} have become ubiquitous  with the growth of recent technologies that are able to generate high-dimensional functional data. 
Although, the vast majority of literature has focused on one-dimensional functional curves \citep{morris2015functional}, recent literature has started investigating models involving more complex types of functional data such as images. For example, neuroimaging analysis using entire brain images as covariates (scalar-on-image regression) for prediction of a continuous outcome have gained in popularity \citep{feng2021brain}. Such approaches are able to discover significantly activated brain voxels and are clearly more attractive compared to methods that evaluate the association between the outcome and each voxel separately \citep{lazar2008statistical}.

Typical scalar-on-image regression approaches need to carefully account for the spatial configuration of hundreds of thousands of voxels, and hence often involve some type of lower dimensional representation for the images such as principal components, wavelet representations, or tensors, along with additional sparsity or shrinkage assumptions designed to tackle the curse of dimensionality. Approaches involving functional principal component analysis (FPCA) \citep{zipunnikov2011functional,feng2019bayesian} often assume that the components driving variability in the images are related to the outcome that may not always be practical, and they are computationally burdensome for high-dimensional images. Moreover, they only use a subset of principal components resulting in information loss that is potentially exacerbated in the presence of noise in images. A limited number of alternate methods involving wavelet-based representations have been proposed for scalar-on-image regression \citep{wang2014regularized,reiss2015wavelet} that provides a desirable avenue to preserve the spatial properties of the image when modeling  regression coefficients. Tensor-based representations for regression coefficients have also been proposed \citep{feng2021brain}, which massively reduce the number of parameters needed to be estimated in the model and are shown to possess desirable asymptotic properties.  However, the finite sample properties of the existing tensor-based approaches are not well understood, particularly for high-dimensional applications where the tensor decomposition may not provide an adequate characterization.


Unfortunately in spite of the growing practical interest, there is negligible development of methods for joint learning of multiple scalar-on-image regression models corresponding to inter-related high-dimensional imaging datasets.  Our joint learning goals are motivated by an increasing interest in data fusion techniques in medical imaging \citep{lahat2015multimodal}, which may involve data on task and rest experiments, or longitudinal neuroimaging data collected in mental health studies such as the Alzheimer's Disease Neuroimaging Initiative (ADNI) \citep{weiner2015introduction}, among others. A joint analysis of such inter-related datasets can leverage common threads of information across experiments or visits that is expected to lead to greater predictive accuracy and higher power to detect true signals, and produce reliable estimates that are biologically interpretable \citep{kundu2019novel}. One can potentially leverage multi-task learning methods in  machine learning literature \citep{zhang2018overview,tang2016fused,li2014meta,lounici2011oracle} and related methods for our purposes. Unfortunately, these existing approaches are not designed to tackle  integrative analysis involving high dimensional images with spatially distributed voxels, and their theoretical and numerical properties are have not been investigated in the presence of noisy functional covariates. Our extensive numerical studies and ADNI analysis reveal that the presence of noise may potentially mask the common patterns across inter-related images, which consequently hinders the ability of existing multi-task learning approaches to learn such patterns and eventually results in poor performance.

The presence of noise in brain images is not surprising, given that measurement errors are likely to arise due to technological limitations, operator performance, equipment, environment, and other factors  \citep{vaishali2015review}. Although standard pre-processing steps are applied to neuroimaging data prior to analysis, they are not expected to completely alleviate the noise in these images. Unfortunately, existing neuroimaging studies do not account for noise in pre-processed images, which is consistent with the predominant practices in biomedical studies. 
Inadequate noise correction can result in estimation bias in the direction of zero that is known as attenuation to the null \citep{carroll1994measurement}. This phenomenon is also clearly evident for our ADNI analysis (Section \ref{sec:adni}) where standard approaches without noise-correction discover negligible brain activations. We note that standard denoising steps in scalar-on-function regression approaches  \citep{ramsay2005functional} may not be biologically meaningful for our neuroimaging applications that already involve a very specific set of pre-processing steps for the images, and they induce additional computational burden. 




Although there is a rich literature on measurement error models with scalar covariates \citep{carroll1994measurement}, there is (unfortunately) a limited literature on scalar-on-function regression with noisy functional predictors, which can not be directly adapted to our settings of interest involving multi-task learning with high-dimensional noisy images.  Such approaches often rely on corrected least squares (OLS) estimators that account for bias due to the presence of noise \citep{crambes2009smoothing} that  may not be applicable to settings when the model dimensions increase much faster than the sample size without additional regularization, or add an additional level of hierarchy by assigning a probability model on the observed functional covariates \citep{james2002generalized, goldsmith2011vb}. These limited approaches have not been adapted to applications with high-dimensional noisy images involving unknown error variances, and their finite sample theoretical properties remain unclear in such settings. In addition, it is not evident whether these methods for noisy functional predictors can be directly applied to multi-task learning problems with an added goal of ensuring model parsimony. Alternative Monte Carlo simulation based approaches such as simulation-extrapolation (SIMEX)  \citep{cook1994simulation} that were originally designed for univariate or lower dimensional covariates, also suffer from similar drawbacks.

In this article, we develop a fundamentally novel approach for the joint analysis of multiple scalar-on-image regression models with high-dimensional noisy images that uses wavelet expansions and grouped penalties for sparse multi-task learning. In particular, we propose a corrected M-estimation approach that adjusts for the bias arising due to noisy images by projecting the solution onto a space of admissible solutions. The proposed approach uses minimal assumptions that involve sub-Gaussian distributions on the true image and additive noise terms with unstructured covariance structures. In order to tackle the curse of dimensionality arising due to high-dimensional images and to enable multi-task learning, we employ grouped penalties such as the non-convex group bridge \citep{huang2009group} as well as the convex $L_{1,q}$ penalty, on the functional regression coefficients. The group bridge penalty promotes differential sparsity patterns across datasets, whereas the group lasso penalty encourages more similar sparsity patterns designed for robust learning across datasets. Since a closed form solution under the corrected optimization criteria is challenging, we propose a computationally efficient projected gradient descent algorithm that approximates the optimal solution of the model parameters in the presence of noisy images. The proposed approaches translate to locally sparse brain activations, i.e. functional regression coefficients that are zero or non-zero over spatially contiguous regions, which adhere to the biological reality of  locally concentrated brain activations in our motivating neuroimaging applications. 

We establish attractive theoretical justifications for the proposed methods for high-dimensional applications where the number of voxels ($p$) increases exponentially with sample size ($n$) for all the $M$ inter-related imaging datasets. Beginning with the case without measurement error in images, we establish weak oracle properties under the group bridge penalty, and we justify the choice of the $L_{1,q}$ penalty by appealing to the desirable theoretical properties that has already been established in literature in the case of covariates without measurement error \citep{lounici2011oracle,negahban2012unified}.  Moving on to the case with images having voxel-specific additive errors, we derive finite sample statistical error bounds for the optimal solutions explicitly in terms of $(n,p,M)$, for both non-convex and convex grouped penalty functions, which become vanishingly small with high probability as $n$ grows to infinity. In addition, we derive finite sample optimization error bounds which illustrate that the iterations of the projected gradient descent under the convex grouped penalties converges with high probability to the optimum solution, which ensures the legitimacy of the computed parameter estimates. Extensive numerical studies conclusively illustrate the gains under the proposed methods over competing multi-task learning approaches without noise correction as well as noise corrected scalar-on-image regression without multi-task learning, in terms of recovery of true signals and predictive performance. We apply the proposed methods to analyze the longitudinal ADNI brain MRI images, which illustrate the predictive gains under the proposed approach when modeling cognitive outcomes, and provides clear evidence regarding the ability of the proposed noise corrected multi-task learning method to detect biologically meaningful brain activations. In contrast, other multi-task learning methods without noise correction result in poor prediction, and negligible or absent brain activations that is consistent with the attentuation to the null phenomenon in literature.

This article makes several significantly novel contributions. First, to our knowledge, the proposed approach is one of the first methods for integrative analysis of multiple scalar-on-image regressions involving inter-related high-dimensional noisy images that provides significant practical gains over existing methods. Hence this approach expands the literature on scalar-on-image regression models without measurement error to scenarios involving multi-task learning in the presence of noisy images. Second, we derive finite sample error bounds for the model parameters explicitly in terms of $(n,p,M)$, which is one of the first such results under grouped non-convex and convex penalties involving noisy functional covariates. Such results provide non-trivial generalizations of previous results in \cite{loh2012} who focused on linear regression under $L_1$ penalties involving noisy scalar covariates without multi-task learning. We note that deriving such error bounds  is not straightforward due to the inherent non-convexity in the loss function that results from the presence of noise (see Section 3). Third, we derive optimization error bounds under the computationally efficient projected gradient descent algorithm to approximate the optimal solution under the grouped $L_{1,q}$ penalty, which guarantees that the computed parameter estimates are well-behaved. To our knowledge, this is one of the first such results involving noisy functional predictors and under grouped penalties, and is motivated by developments in \cite{agarwal2012}.

 Section \ref{sec:method_noME} develops the joint scalar-on-image regression approach and theory corresponding uncorrupted images, while Section \ref{sec:method_ME} extends this approach for noisy images. Section \ref{sec:sims} involves extensive simulation studies, Section \ref{sec:adni} applies the methods to ADNI data, and Section \ref{sec:discuss} contains further discussions. Supplementary Materials contain additional materials. 

\section{Multi-task learning without Measurement Errors}
\label{sec:method_noME}
Our goal is to propose an approach for joint learning for multiple scalar-on-surface regressions.  Denote the scalar continuous outcome $y\in \Re$ that is regressed on an image $X$ defined over a $d$-dimensional surface and  observed at a discrete set of voxels $\{{\bf v}_1,\ldots,{\bf v}_p \}$. Here  ${\bf v}_l \in [0,1 ]^d$ without loss of generality, where $d=2$ or $3$ in practice corresponding to two- or three-dimensional (2-D or 3-D) images. Moreover, the images for all the subjects and data sources are registered to a common template that is standard in medical imaging applications \citep{avants2011reproducible}. 
For the purposes of exposition and illustration, we will consider the situation with 2-D images as functional predictors. However, our framework is naturally applicable to 3-D images (see Supplementary Materials). Further, it is straightforward to include additional datasource-specific scalar covariates in the modeling framework, but they are omitted in the following discussions in order to preserve simplicity of notations. Let $y_{mi}$ and $X_{mi}$ denote the outcome and the imaging predictor for subject $i$ $(i=1,\cdots,n_m)$ from data source $m$ $(m=1,\cdots,M)$, where $x_{mi}({\bf v}_k)$ denotes the observed MRI image at voxel ${\bf v}_k$ corresponding to $X_{mi}$. We assume that $|\int X_{mi}({\bf v})\mathrm{d}{\bf v}|<\infty$, which is reasonable when ${\bf v} \in [0,1 ]^d$. The scalar-on-image regression model for the $m$th data source can be written as:
\begin{equation}
 y_{mi} = \beta_{m0} + \int  X_{mi}({\bf v})\beta_m({\bf v})\mathrm{d}{\bf v} + \epsilon_{mi}, \mbox{ } \epsilon_{mi}\stackrel{i.i.d.}{\sim} N(0,\sigma_m^2), \mbox{ } m=1,\cdots,M; i=1,\cdots,n_m,
\label{eqt:linear_model}
\end{equation}
where $\epsilon_{mi}$ denotes the random error term that is assumed to be normally distributed with data source-specific residual variance, and $\beta_m(\cdot)$ is the functional regression coefficient that captures the effects of the functional predictor on the outcome corresponding to the $m$th data source. These functional regression coefficients are estimated jointly across $M$ datasets under an integrative learning framework, as elaborated in the sequel. We allow the number of subjects to vary across the data sources, which gives us flexibility in handling missing data at follow-up visits that is encountered in our motivating ADNI study.

We propose a wavelet-based decomposition for the 2-D  images that provides a multiscale representation to accommodate varying degrees of smoothness \citep{reiss2015wavelet} as:
\begin{equation}
    x_{mi}({\bf v}) = \sum_{k,l=0}^{2^{j_0}-1} c_{mi,j_0,\{k,l\}}^0\phi_{j_0,\{k,l\}}({\bf v}) + \sum_{j=j_0}^J\sum_{k,l=0}^{2^j-1}\sum_{q=1}^3 c_{mi,j,\{k,l\}}^q\psi_{j,\{k,l\}}^q({\bf v})
\label{eqt:image_decomp}
\end{equation}
where $j_0$ is the primary level of decomposition that controls the number of basis elements in the multi-scale representation, $J$ denotes the maximum level of decomposition, and  $\{\phi_{j_0,\{k,l\}},k,l=1,\ldots,2^{j_0}-1\}$ and $\{\psi_{j,\{k,l\}}^q, j=j_0,\cdots,J,k,l=0,\ldots,2^j-1,q=1,\cdots,3 \}$ denote pairwise orthonormal wavelets. 
The wavelet basis functions in (\ref{eqt:image_decomp}) can also be expressed as $\phi_{j_0,\{k,l\}}({\bf v}) = \phi_{j_0,k}(v_1)\phi_{j_0,l}(v_2), \mbox{ } \psi_{j,\{k,l\}}^1({\bf v}) = \psi_{j,k}(v_1)\phi_{j,l}(v_2)$, $\psi_{j,\{k,l\}}^2({\bf v}) = \phi_{j,k}(v_1)\psi_{j,l}(v_2)$, $\psi_{j,\{k,l\}}^3({\bf v}) = \psi_{j,k}(v_1)\psi_{j,l}(v_2),$
where ${\bf v}=(v_1,v_2)$ and $\phi_{j,\cdot}(\cdot), \psi_{j,\cdot}(\cdot)$ are the one-dimensional father and mother wavelets of level $j$. Using orthonormality, the wavelet coefficients in (\ref{eqt:image_decomp}) can be calculated by
$c_{mi,j_0,\{k,l\}}^0 = \langle {\bf x}_{mi},\phi_{j_0,\{k,l\}}\rangle$ and $c_{mi,j,\{k,l\}}^q = \langle {\bf x}_{mi},\psi_{j,\{k,l\}}^q\rangle$, where we use  $\langle f_1, f_2 \rangle = \int f_1({\bf v})f_2({\bf v}) d{\bf v}$ to denote the inner product and ${\bf x}_{mi}$ denotes the vectorized image. Due to discrete wavelet transform, we usually require the observed number of locations along all dimensions to be the same (power of 2). If the original functional data does not fulfill this requirement, we can easily pad zero values around it and increase the dimension to the nearest high power of 2 \citep{reiss2015wavelet}, which we denote as $p_0$. Then the maximum level $J=\log_2(p_0)-1$, and $p=p_0^2$ is also the total number of wavelet coefficients in (\ref{eqt:image_decomp}).

 
 We assume the true functional regression coefficients to have bounded total variation, i.e. $\|\beta^0_m({\bf v})\|_V = \int_0^1\int_0^1 \big|\nabla \beta^0_m({\bf v})\big|\mathrm{d}{\bf v}< \infty$, and bounded amplitude, where  $\nabla$ denotes the partial derivative in the general sense of distributions \citep{ziemer2012weakly}. This implies that the true regression coefficients lie in the space of functions $
\mathcal{T} := \big\{\beta({\bf v})\in L^2[0,1]^2: \|\beta({\bf v})\|_V<\infty, \big\|\beta({\bf v})\big\|_\infty<\infty\big\}$ .  We can express $\beta_m^0\in \mathcal{T}$ as 
$
    \beta_m^0({\bf v}) = \sum_{k,l=0}^{2^{j_0}-1} a^{0}_{m,j_0,\{k,l\}}\phi_{j_0,\{k,l\}}({\bf v}) + \sum_{j=j_0}^\infty \sum_{k,l=0}^{2^j-1}\sum_{q=1}^3 d_{m,j,\{k,l\}}^{0q}\psi_{j,\{k,l\}}^q({\bf v})=B^T({\bf v})\bm{\eta}_m^0 + e_m^0
$
for a separable, compactly supported and orthonormal wavelet basis of $L^2[0,1]^2$, where  $\{ a^{0}_{m,j_0,\{k,l\}}\}$ and $\{d_{m,j,\{k,l\}}^{0q} \}$ are the true scaling and dilation coefficients, the primary decomposition level is assumed to be known at $j_0$, $\bm{\eta}_m^0$ is corresponds to the first $p$ coefficients of the true wavelet coefficient vector, and $e_m^0(\cdot)$ is the approximation term such that $\|e_m^0(\cdot) \|_\infty = O(2^{-J})$=$O(p^{-1/2})$ based on Theorem 9.18 in \cite{mallat1999wavelet}. Similar specifications can be found in \cite{wang2017generalized}. Since in practice, the functional data are only observed at discrete locations, and given that $ e_m^0(\cdot)$ rapidly decreases to zero for large $p$, we will focus on recovering the truncated coefficient vector $\bm{\eta}_m^0$ and we will view it as the true wavelet coefficients for our subsequent discussions.

In the above spirit, we will use a finite basis expansion for model fitting. In particular, we use $
    \beta_m({\bf v}) = \sum_{k,l=0}^{2^{j_0}-1} a_{m,j_0,\{k,l\}}\phi_{j_0,\{k,l\}}({\bf v}) + \sum_{j=j_0}^J \sum_{k,l=0}^{2^j-1}\sum_{q=1}^3 d_{m,j,\{k,l\}}^q\psi_{j,\{k,l\}}^q({\bf v})
$, where the unknown coefficients 
$ a_{m,j_0,\{k,l\}} = \langle \beta_m,\phi_{j_0,\{k,l\}}\rangle$ and $d_{m,j,\{k,l\}}^q = \langle\beta_m,\psi_{j,\{k,l\}}^q\rangle
$ can be directly computed from the images for a given choice of basis functions. The wavelet representation is flexible enough to allow the functional regression coefficient to be estimated at different levels of smoothness via different levels of $j_0$. One can rewrite the model (\ref{eqt:linear_model}) as:
\begin{equation}
    y_{mi} = \beta_{m0} + \bm{c}_{mi}^T{\bm\eta}_m + \epsilon_{mi},  \mbox{ } \epsilon_{mi}\stackrel{i.i.d.}{\sim} N(0,\sigma_m^2), \mbox{ } i=1,\ldots,n_m, \mbox{ } m=1,\ldots,M,
\label{eqt:model_rewrite}
\end{equation}
where $\small \bm{c}_{mi}^T{\bm\eta}_m= \sum_{k,l=0}^{2^{j_0}-1} c_{mi,j_0,\{k,l\}}^0 a_{m,j_0,\{k,l\}} + \sum_{j=j_0}^J\sum_{k,l=0}^{2^j-1}\sum_{q=1}^3 c_{mi,j,\{k,l\}}^q d_{m,j,\{k,l\}}^q$ is the linear mean term, ${\bm\eta}_m= (\eta_{m1},\ldots,\eta_{mp})'$ denotes the vector of unknown wavelet coefficients corresponding to $\beta_m({\bf v})$ 
, and $\bm{c}_{mi} (p\times 1)$  denotes the collection of coefficients corresponding to the decomposition of $X_{mi}$ in (\ref{eqt:image_decomp}) that can be computed explicitly. Model (\ref{eqt:model_rewrite}) is now a standard linear regression model with known design matrix $\bm{C}_m = (\bm{c}_{m1},\cdots,\bm{c}_{mn_m})^T$  and unknown wavelet coefficients ${\bm \eta}_m, m=1,\cdots,M$, corresponding to the $m$th data source, which are jointly estimated across datasets (as elaborated in the sequel) and can be used to reconstruct the functional regression coefficients. We note that by location transformation, we can assume $\beta_{m0} = 0, m=1,\cdots,M$, without loss of generality, and hence we will ignore the intercept terms in the following discussions. Model (\ref{eqt:model_rewrite}) represents a discretized version of the original functional linear model in (\ref{eqt:linear_model}) that will be used throughout this article. We note that in contrast to the working model (\ref{eqt:model_rewrite}), the true model is given as 
$y_{mi} = \int X_{mi}({\bf v})\beta_m^0({\bf v})\rm{d}{\bf v} + \epsilon_{mi}
       = \bm{c}_{mi}^T\bm{\eta}_m^0 + \int X_{mi}({\bf v})\bm{e}_m^0({\bf v})\rm{d}{\bf v} + \epsilon_{mi}$, using the above discussions.


One can use different types of wavelet bases in (\ref{eqt:image_decomp})--(\ref{eqt:model_rewrite}) - see \citep{walker2008primer} for more details on different choices of wavelet basis. One possible choice is the Haar wavelets  \citep{wang2014regularized, wang2017generalized} that  results in piecewise constant approximations of the signal due to one vanishing moment. The Haar wavelets can be generalized to accommodate higher number of vanishing moments via the Daubechies wavelets \citep{reiss2015wavelet}, which is able to capture diverse types of signals while preserving model parsimony. The choice of the wavelet bases can be tuned to the particular application, as needed.



In order to ensure sparsity in the estimated coefficients that reflects the biological reality of a small subset of activated brain locations driving the outcome, suitable grouped penalty functions $\rho(\cdot)$ are imposed that facilitate joint learning. In particular, we propose to solve the optimization problem:
$ \max_{\bm{\eta}} \big\{-\frac{1}{2}\sum_{m=1}^M \sum_{i=1}^{n_m} \Big(y_{mi} - \langle\bm{c}_{mi},\bm{\eta}_m\rangle\Big)^2 - \lambda\rho(\bm{\eta})\big\}$, where $\rho(\bm{\eta})$ may correspond to convex penalty functions such as the $L_{1,q}$ penalty $\rho(\bm{\eta}) = \sum_{j=1}^p\Big(\sum_{m=1}^M|\eta_{mj}|^q\Big)^{1/q} (q>1)$, that includes the group lasso when $q=2$, as well as non-convex penalties such as the group bridge, i.e. $\rho(\bm{\eta}) = \sum_{j=1}^p\Big(\sum_{m=1}^M|\eta_{mj}|\Big)^{1/2}$.  The convex and non-convex penalties lead to different modes of joint learning by promoting varying sparsity patterns. The group lasso penalty is expected to work better in cases with greater homogeneity across datasets, while the group bridge penalty is recommended for scenarios with more heterogeneous data sources. The choice of these penalties is motivated by our primary goal of data fusion and associated theoretical properties that are already established in literature for the case without measurement error. For example, consistency properties under the group lasso penalty \citep{nardi2008asymptotic,lounici2011oracle} are well-known, while the asymptotic properties \citep{huang2009group} and weak oracle properties for binary outcomes \citep{li2014meta} of the group bridge penalty have been established in literature.

{\noindent \emph{Weak Oracle properties under group bridge with uncorrupted images:}} We will establish weak oracle properties \citep{lv2009unified} under group bridge that will extend the results in \cite{li2014meta} corresponding to binary outcomes involving scalar covariates to the case of scalar-on-image regression with continuous outcomes. For the ease of notation and without loss of generality, we assume that all $M$ data sources have $n$ samples in the following discussion. However, the proposed methodology and theoretical developments are equally applicable for unequal sample sizes across data sources. One can rewrite the optimization problem as
\begin{equation}
    \max_{\bm{\eta}}\bigg\{\bm{y}^T\bm{C}\bm{\eta} - \frac{1}{2}\bm{\eta}^T\bm{C}^T\bm{C}\bm{\eta}-n\lambda_n\rho(\bm{\eta})\bigg\}
\label{eqt:max4}
\end{equation}
where $\rho(\bm{\eta}) = \sum_{j=1}^p\Big(\sum_{m=1}^M|\eta_{mj}|\Big)^{1/2}$, and $\bm{C}_{(Mn\times Mp)}$ is a block-diagonal design matrix whose $m$th block $\bm{C}_m$ corresponds to known wavelet coefficients from the $m$th data source. 
Consider the following partition of the index set $\{1,\ldots,p \}$: $I=\{(m,j)|\eta_{mj}^{0} \neq 0, \bm{\eta}_{(j)}^{0} \neq \mathbf{0}\}$, $II=\{(m,j)|\eta_{mj}^{0} = 0, \bm{\eta}_{(j)}^{0} \neq \mathbf{0}\}$ and $III=\{(m,j)| \bm{\eta}_{(j)}^{0} = \mathbf{0}\}$ where $\bm{\eta}_{(j)}^{0}=(\eta_{1j}^0,\cdots,\eta_{Mj}^0)^T$ denotes the true wavelet coefficients corresponding to the $j$-th wavelet basis function. Set $I$ denotes the indices for the true nonzero coefficients across all data sources, set $II$ denotes the indices for those true wavelet coefficients that are zero for some data sources but not others, while set $III$ denotes the indices for those wavelet coefficients that are zero across all data sources. It is clear that the three sets are mutually exclusive. Further, let $s=|I|$ be the true sparsity level, let $d = 0.5\min\{|\eta_{mj}^0|:\eta_{mj}^0\in I\}$ with $d/\log n\asymp n^{-\alpha_d}$ where $\alpha_d\le\gamma,  \gamma\in(0,1/2]$, and denote $l=\min_{\{j:\bm{\eta}_{(j)}^0\ne\bm{0}\}}\|\bm{\eta}_{(j)}^0\|_1^{1/2}$, $L=\max_{\{j:\bm{\eta}_{(j)}^0\ne\bm{0}\}}\|\bm{\eta}_{(j)}^0\|_1^{1/2}$. Define the neighborhood  $\mathcal{N}_0 = \{\bm{\delta}\in\mathbb{R}^s:\|\bm{\delta}-\bm{\eta}_I^0\|_\infty\le d\}$, and the constant  $\kappa_0=\max_{\bm{\delta}\in\mathcal{N}_0}\max_{\{j|\bm{\delta}_{(j)}\ne\bm{0}\}}4^{-1}\|\bm{\delta}_{(j)}\|_1^{-3/2}$, where  $\bm{\delta}_{(j)}=(\delta_{1j},\cdots,\delta_{Mj})^T$ with $\delta_{mj}=0$ for $(m,j)\notin I$. 

Consider the standardized design matrix $\bm{C}$ denoted as $\Tilde{\bm{C}}$, such that $\|\Tilde{\bm{c}}_{mj}\|_2=\sqrt{n}$. We denote $\|\cdot\|_\infty$ as the supremum norm, and denote $\mathcal{O}$ and $o$ as the big-O and little-o notations. Also $a\asymp b$ implies $a,b,$ are on the same order. We will assume  the following conditions:

{\noindent 
{\it (C1)} $ \alpha_p=\min(1/2,2\gamma-\alpha_s)$, where $s\asymp n^{\alpha_s}, \alpha_s<1, \log p\asymp n^{1-2\alpha_p}$ and $\gamma\in(0,1/2]$;\\
 {\it (C2)} $\|(\Tilde{\bm{C}}_I^T\Tilde{\bm{C}}_I)^{-1}\|_\infty = \mathcal{O}(b_sn^{-1})$, $b_s = o(n^{1/2-\gamma}\sqrt{\log n})$;
{\it (C3)} $\|\Tilde{\bm{C}}_{II}^T\Tilde{\bm{C}}_I(\Tilde{\bm{C}}_I^T\Tilde{\bm{C}}_I)^{-1}\|_\infty \le l/(2L)$.\\
}
Assumption {\it (C1)} places conditions on the rate of growth for the true sparsity level ($s$) and allows $p$ to grow much faster than $n$. {\it (C2)} essentially requires $\frac{1}{n}\Tilde{\bm{C}}_I^T\Tilde{\bm{C}}_I$ to be non-singular and that the supremum norm of $(\Tilde{\bm{C}}_I^T\Tilde{\bm{C}}_I)^{-1}$ has a lower bound as in equation (15) in \cite{fan2011nonconcave}, while {\it (C3)} is similar to the irrepresentability condition in literature  \citep{zhao2006model}. Given {\it (C1)-(C3)}, Theorem \ref{theo:thm1} formalizes the weak oracle property. 
\begin{theorem}
Suppose the conditions (C1)-(C3) hold. For $\lambda_n$ satisfying $\lambda_{n} \asymp n^{-\alpha_{\lambda}}$ with  $\alpha_{\lambda} < \alpha_{p}$, $\lambda_n b_s = o(n^{-\alpha_d/2-\gamma}\log n)$ and $\lambda_n \kappa_0 = o(\tau_0)$, where  $\tau_0=\lambda_{\min}(n^{-1}\Tilde{\bm{C}}_I^T\Tilde{\bm{C}}_I)$, there exists a local maximizer $\bm{\hat{\eta}}$ of (\ref{eqt:max4}), such that: (a) $\bm{\hat{\eta}}_{II \cup III}=\bm{0}$; and (b) $\|\hat{\bm{\eta}}_{I}-\bm{\eta}_{I}^{0}\|_{\infty} \leq n^{-\gamma}\log n$, with probability greater than $1-2\{sn^{-1}+(Mp-s)e^{-n^{1-2\alpha_p}\log n}\}$ for sufficiently large $n$.
\label{theo:thm1}
\end{theorem}
Property (a) indicates that the oracle estimator for the truly zero wavelet coefficients are  estimated correctly with high probability tending to one as $n\to\infty$. Property (b) indicates that with high probability tending to one as $n$ increases, the error under the oracle estimator (in terms of the supremum norm) corresponding to the truly nonzero wavelet coefficients is bounded by a term that goes to zero. Together properties (a) and (b) in Theorem \ref{theo:thm1} imply the weak oracle property, which hold for certain strict local maximizers that satisfies the KKT conditions as detailed in Lemma 1 in the Supplementary Materials. The results for the wavelet coefficients in Theorem \ref{theo:thm1} can be used to deduce the non-asymptotic error bounds for the corresponding oracle estimators of the functional regression coefficients $\{\hat{\bm{\beta}}_1(\cdot),\ldots,\hat{\bm{\beta}}_M(\cdot)\}$ and the predicted means, as captured via the following result.


\begin{corollary}
If Theorem \ref{theo:thm1} holds, then $\big|\hat{\beta}_m({\bf v}) - \beta_m^0({\bf v})\big| \leq \tau_m({\bf v})sn^{-\gamma}\log n + O(p^{-1/2})$ and 
$\Big|\int\bm{X}_{mi}({\bf v})\hat{\beta}_m({\bf v})\mathrm{d}{\bf v} - \int\bm{X}_{mi}({\bf v})\beta_m^0({\bf v})\mathrm{d}{\bf v}\Big| \le \iota_m sn^{-\gamma}\log n + O(p^{-1/2})$, 
for all $m \in \{1,\cdots,M\}$, 
where $\tau_m({\bf v})$ and $\iota_m $ can be calculated from the images.
\label{theo:lem2}
\end{corollary}

\section{Multi-task learning with Measurement Errors}
\label{sec:method_ME}
 We now generalize our multi-task learning scalar-on-image regression approach to the case of images with measurement errors that is the main focus of this article. We assume an additive measurement error model, i.e. $ z_{mi}({\bf v}) = x_{mi}({\bf v}) + u_{mi}({\bf v}),\mbox{ } i=1,\cdots,m_n,$ $m=1,\cdots,M,$
where $x_{mi}({\bf v})$ denotes the true unobserved image at voxel ${\bf v}$, while $z_{mi}({\bf v})$ denotes the observed noisy image with the measurement error $u_{mi}({\bf v})$. While we still assume the true model to be $y_{mi} = \beta_{m0} + \int  X_{mi}({\bf v})\beta_m({\bf v})\mathrm{d}{\bf v} + \epsilon_{mi}$, we relax the distribution of the random error $\epsilon_{mi}$ to be sub-Gaussian with parameter $\sigma_m^2$. Moreover, the working model used for fitting the data would replace the true image $x_{mi}({\bf v})$ (that is unobserved) by its noisy counterpart $ z_{mi}({\bf v})$ in (\ref{eqt:image_decomp}), which results in a different M-estimation criteria compared to  (\ref{eqt:max4}) (see equation (\ref{eqt:corrected})). We capture the randomness of the noisy images by assuming that the vectorized true image $\bm{x}_{mi}=\big(x_{mi}({\bf v}_1),\ldots, x_{mi}({\bf v}_p)\big)$ and the measurement errors $\bm{u}_{mi}=\big(u_{mi}({\bf v}_1),\ldots, u_{mi}({\bf v}_p)\big)$ are independently distributed as sub-Gaussian random variables (see Supplemental Materials) with parameters $(\Sigma_m^x,\sigma_x^2)$ and $(\Sigma_u,\sigma_u^2)$ respectively, which ensures bounded tails.

Let us denote the matrix of wavelet basis corresponding to the $p$ discrete locations as $B = (\bm{b}_1,\cdots,\bm{b}_p)$ where $\bm{b}_j = \big(b_j({\bf v}_1),\cdots,b_j({\bf v}_p)\big)^T$ denotes the realizations of the $j$-th basis function at all $p$ discrete locations ($j=1,\cdots,p$), so that $B^T B=I$. We define $\bm{w}_{mi}=B^T \bm{z}_{mi} = B^T\bm{x}_{mi} + B^T\bm{u}_{mi} = \bm{c}_{mi} + B^T\bm{u}_{mi}$ as the inner product between the noisy observed image and the wavelet basis functions over all voxels. Hence, $\bm{w}_{mi}$ can be considered as an adaptation of the image wavelet coefficients $\bm{c}_{mi}$ in (\ref{eqt:image_decomp}) to the case of noisy images. 
A naive solution that ignores the noise in the images would be to replace $\bm{c}_{mi} $ with $\bm{w}_{mi}$ in the scalar-on-image regression model  (\ref{eqt:model_rewrite}), but  this strategy would result in inconsistent estimation under the criteria (\ref{eqt:max4}) \citep{sorensen2015measurement}. A potential solution to remedy the problem that is motivated by \cite{loh2012}, is to use a corrected M-estimator that resembles (\ref{eqt:max4}) but adjusts for the additive noise. In this context, note that the matrix $\bm{C}_m^T \bm{C}_m$ in (\ref{eqt:max4}) corresponding to the unobserved true images can be approximated by $\hat{\bm{\Gamma}}_m=\frac{1}{n}\bm{W}_m^T\bm{W}_m-B^T\bm{\Sigma}_uB$, via the relationship $\bm{w}_{mi}=B^T \bm{z}_{mi}=  \bm{c}_{mi} + B^T\bm{u}_{mi}$, where $\bm{W}_m = (\bm{w}_{m1},\cdots,\bm{w}_{mn})^T$. We propose the following noise corrected version of (\ref{eqt:max4}) as  
 \begin{eqnarray}
   \underset{\rho(\bm{\eta})\le R}{\min} \bigg[\sum_{m=1}^M\bigg\{\frac{1}{2}\bm{\eta}_m^T\hat{\bm{\Gamma}}_m\bm{\eta}_m - \langle\hat{\bm{\gamma}}_m,\bm{\eta}_m\rangle\bigg\} + \lambda_n\rho(\bm{\eta})\bigg], 
\label{eqt:corrected}
\end{eqnarray}
where $\hat{\bm{\gamma}}_m=\frac{1}{n}\bm{W}_m^T\bm{y}_m$, $\bm{y}_m=(y_{m1},\cdots,y_{mn})^T$,  $\rho(\bm{\eta})$ denotes grouped penalty functions for multi-task learning, and the remaining terms other than  $\rho(\bm{\eta})$ in (\ref{eqt:corrected}) represents the loss function $\mathcal{L}$ that makes use of a corrected variance term $\hat{\bm{\Gamma}}_m$ and a cross-product $\hat{\bm\gamma}_m$. Lemma \ref{thm:deviation} illustrates that $(\hat{\bm{\Gamma}}_m,\hat{\bm\gamma}_m)$ serve as surrogates for $[Var({\bf c}_{mi})]$ and $[Var({\bf c}_{mi})]{\bm \eta}^{0}_m$ respectively, where approximation error is shown to decrease to zero as $n\to \infty$ even when $p>>n$. This property indicates the resemblance between the corrected criteria (\ref{eqt:corrected}) and criteria (\ref{eqt:max4}) corresponding to no measurement error, since the terms $C^T{\bf y}$ and $C^TC$ in (\ref{eqt:max4}) are also unbiased estimators for $[Var({\bf c}_{mi})]{\bm \eta}^{0}_m$ and $Var({\bf c}_{mi})$ respectively, in the absence of noise. 

\begin{lemma} (deviation condition) The surrogates  $(\hat{\bm{\Gamma}}_m,\hat{\bm{\gamma}}_m)$ satisfy 
(i) $\small\Big\|\hat{\bm{\gamma}}_m-B^T\bm{\Sigma}_m^x B\bm{\eta}^0_m\Big\|_\infty\le \phi\sqrt{\frac{\log p}{n}}$; and (ii)  $\small \Big\|\Big(\hat{\bm{\Gamma}}_m-B^T\bm{\Sigma}_m^x B\Big)\bm{\eta}^0_m\Big\|_\infty\le \phi\sqrt{\frac{\log p}{n}}$, with probability at least $1-c_1\exp\{-c_2\log p\}$, where $\sigma^2 = \sigma_x^2+\sigma_u^2$, $\phi = \max_{m}\big\{c_0\sigma(\sigma_m + \sigma\|\bm{\eta}_m^0\|_1)\big\} + c_0^\prime\sigma p^{-1/2}$, and constants $c_0,c_0^\prime,c_1,c_2>0$.
\label{thm:deviation}
\end{lemma}

Another unique feature of (\ref{eqt:corrected}) is that it requires the solution to be restricted to the ball $\rho(\bm{\eta})\le R$, defined in terms of the penalty $\rho(\cdot)$. This  restriction is imposed to ensure the stability of solutions, since  $\hat{\bm{\Gamma}}_m$'s are generally not positive semi-definite in the presence of noise, making the  loss $\mathcal{L}$  in (\ref{eqt:corrected}) non-convex. The restricted solution enables one to tackle a potentially large number of negative eigen values in $\hat{\bm{\Gamma}}_m$ due to noisy images, which could otherwise lead to the objective function $\mathcal{L}$  being unbounded from below in an extreme case. In what follows, we will first use the assumption that the covariance matrix for the measurement error $\bm{\Sigma}_u$ is known to develop our model and theoretical properties, and subsequently relax this assumption and generalize the results to unknown noise covariance matrices that are empirically estimated. Lemma \ref{thm:deviation} and the above discussions provide an intuition regarding the proposed corrected criteria in (\ref{eqt:corrected}), which is motivated by \cite{loh2012}. 

\vskip 12pt


{\noindent \bf \large 3.1. Theoretical properties under noisy images}  

In order to establish theoretical properties corresponding to images with measurement errors under criteria (\ref{eqt:corrected}), it is necessary to characterize the behavior of the matrix $\hat{\bm{\Gamma}}$ (referring to any of $\hat{\bm{\Gamma}}_1,\cdots,\hat{\bm{\Gamma}}_M$) via some lower restricted eigen value (lower-RE) conditions that places lower bounds on quadratic terms $\bm{\eta}^T\hat{\bm{\Gamma}}\bm{\eta}$ in  (\ref{eqt:corrected}). Such conditions prevent the objective function from being unbounded from below when there are a large number of negative eigen values for $\hat{\bm{\Gamma}}$  in the presence of noise. Also, the lower-RE condition ensures that the curvature is not overly flat, since $\hat{\bm{\Gamma}}$ represents the curvature of the loss function (equivalent to a Hessian matrix in classical literature). Sufficiently curved loss functions are needed to ensure that optimum solutions are able to converge  sufficiently close to the true parameter values, given that a small loss difference $|\mathcal{L}(\hat{\bm{\eta}})-\mathcal{L}(\bm{\eta}^0)|$ will translate to a small error $|\hat{\bm{\eta}}-\bm{\eta}^0|$. To this effect, we define the following lower-RE condition that is similar to those that have been used extensively in penalized regression literature \citep{van2009conditions}, and can be considered a substitute for global strong convexity that can not be guaranteed when $p>>n$.

\begin{definition}
(Lower-RE condition). The matrix $\hat{\bm{\Gamma}}$ satisfies a lower restricted eigenvalue condition with curvature $\alpha_1>0$ and tolerance $\tau>0$ if
$ \bm{\theta}^T\hat{\bm{\Gamma}}\bm{\theta}\ge \alpha_1\|\bm{\theta}\|_2^2 - \tau\|\bm{\theta}\|_1^2, \forall \bm{\theta}\in\mathbb{R}^{p}.$
\end{definition}

It turns out that the lower-RE condition holds with high probability in the presence of noise, given the sub-Gaussian assumptions on the true images and the additive errors. This is shown by Lemma 2 in the Supplementary Materials under certain choices for $\alpha_1$ and $\tau$, which follows from the results in \cite{loh2012}.




We are now in a position to formally establish the finite sample error bounds corresponding to the optimal estimators obtained under (\ref{eqt:corrected}). We denote $\bm{\eta}^0=(\bm{\eta}^{0T}_1,\cdots,\bm{\eta}^{0T}_M)^T$ as the true wavelet coefficients concatenated over $M$ data sources. Further denote the index set for true  wavelet coefficients that have at least one non-zero signal across data sources as $S = \{j:\|\bm{\eta}_{(j)}^0\|_1\ne 0\}$ where the cardinality of $S$ represents the group sparsity of $\bm{\eta}^0$ (denoted by $k$). Similarly, denote the index set of unimportant wavelet coefficients as $S^C = \{j:\|\bm{\eta}_{(j)}^0\|_1=0\}$, let $l=\min_{j\in S}\|\bm{\eta}_{(j)}^0\|_1^{1/2}$, $h_1 = 1+3l^{-1}R$ and $h_2 = 1+3M^{(q-1)/q}$. The following result establishes the statistical ($L_1$ and $L_2$) error bounds corresponding to the group bridge penalty  explicitly in terms of $n,p,M,R,$ and other parameters.

\begin{theorem}\label{thm:MEbound_gbridge}
(statistical error under group bridge) For any $\bm{\eta}^0$ with group sparsity at most k, the global optimum $\hat{\bm{\eta}}$ of the problem (\ref{eqt:corrected}) under the group bridge penalty $\rho(\cdot)$ satisfies the following error bounds with probability at least $1-c_1\exp\{-c_2\log p \}$ for constants $c_1,c_2>0,$ $R\ge\rho(\bm{\eta}^0)$, $\alpha_1/\tau\ge 2h_1^2Mk$, and $\lambda_n\ge 2\phi\sqrt{\frac{\log p}{n}}\max\{l,R\}$:\\
   $ \|\hat{\bm{\eta}}-\bm{\eta}^0\|_2 \le \frac{8h_1\sqrt{Mk}}{\alpha_1}\max\bigg\{\phi\sqrt{\frac{\log p}{n}},\lambda_nl^{-1} \bigg\}, \mbox{ }  \|\hat{\bm{\eta}}-\bm{\eta}^0\|_1 \le \frac{8h_1^2Mk}{\alpha_1}\max\bigg\{\phi\sqrt{\frac{\log p}{n}},\lambda_nl^{-1} \bigg\}$
\end{theorem}


{\noindent \bf Remark 3.1:} While Theorem \ref{thm:MEbound_gbridge} is stated for a global optimum, we note that the result also holds for any local optimum that satisfies the constraint $\mathcal{L}(\hat{\bm{\eta}})\le \mathcal{L}(\bm{\eta}^0)$ for $\mathcal{L}$ as in (\ref{eqt:corrected}).

\begin{corollary}
 The error bounds in Theorem \ref{thm:MEbound_gbridge} hold when the space of admissible solutions in (\ref{eqt:corrected}) is restricted to a $L_1$ ball $\{\bm{\eta}:\|\bm{\eta}\|_1\le R^2\}$ in (\ref{eqt:corrected}), provided that $\|\bm{\eta}^0\|_1\le R^2$ holds.
\label{thm:coro_L1restriction}
\end{corollary}

We note that Theorem \ref{thm:MEbound_gbridge} guarantees that the bound on the statistical error under a non-convex penalty 
goes to zero even when $p>>n$ and in the presence of measurement errors. In addition, Remark 3.1 suggests the existence of local optima corresponding to (\ref{eqt:corrected}) that come arbitrarily close to the true parameters in terms of bounded statistical errors in Theorem \ref{thm:MEbound_gbridge}. Moreover, Corollary \ref{thm:coro_L1restriction} illustrates that the results in Theorem \ref{thm:MEbound_gbridge} are valid when the space of admissible solutions in (\ref{eqt:corrected}) is modified in terms of a $L_1$ ball, which provides computational benefits when deriving parameter estimates under a projected gradient descent algorithm (see Section 4). Similar to Theorem \ref{thm:MEbound_gbridge}, we now establish finite sample error bounds under the convex $L_{1,q}$ penalty, which includes the group lasso ($q=2$) as a special case. We note that although the $L_{1,q}$ penalty is convex, the issues of non-convexity arising due to noise in the images that are encountered in Theorem \ref{thm:MEbound_gbridge} still persist in this scenario as well. We note that Theorems \ref{thm:MEbound_gbridge}-\ref{thm:MEbound_glasso} provide novel finite sample error bounds that go beyond existing results in literature by accommodating non-convexity arising due to high-dimensional noisy images in the setting of multi-task learning involving non-convex and convex grouped penalties.


\begin{theorem} (statistical error under $L_{1,q}$ penalty) \label{thm:MEbound_glasso}
For any $\bm{\eta}^0$ with group sparsity at most $k$, the global optimum $\hat{\bm{\eta}}$ of the problem (\ref{eqt:corrected}) under the $L_{1,q}$ penalty $\rho(\cdot)$ satisfies the following error bounds with probability at least $1-c_1\exp\{-c_2\log p \}$ for constants $c_1,c_2>0,$ $R\ge\rho(\bm{\eta}^0)$, $\alpha_1/\tau\ge 2h_2^2Mk$, $\lambda_n\ge 2\phi\sqrt{\frac{\log p}{n}}M^{(q-1)/q}$:\\
$
    \|\hat{\bm{\eta}}-\bm{\eta}^0\|_2 \le \frac{8h_2\sqrt{Mk}}{\alpha_1}\max\bigg\{\phi\sqrt{\frac{\log p}{n}},\lambda_n \bigg\}, \mbox{ }  \|\hat{\bm{\eta}}-\bm{\eta}^0\|_1 \le \frac{8h_2^2Mk}{\alpha_1}\max\bigg\{\phi\sqrt{\frac{\log p}{n}},\lambda_n \bigg\}
$
\end{theorem}


\noindent\textbf{Remark 3.2:} Although Theorems \ref{thm:MEbound_gbridge}-\ref{thm:MEbound_glasso} guarantee accurate recovery of the estimated wavelet coefficients, no such finite sample guarantees are available for prediction error. More discussions can be found in \cite{sorensen2015measurement}. However, the proposed methods have good predictive performance as evident from the extensive numerical studies in the sequel.

 \noindent\textbf{Remark 3.3:} The $L_2$ error bounds can be expressed as $ \frac{4}{\tau h_1 \sqrt{Mk}} \max\{\phi\sqrt{\log(p)/n},\lambda_n l^{-1} \}$ under Theorem 3.1, and as $ \frac{4}{\tau h_2 \sqrt{Mk}} \max\{\phi\sqrt{\log(p)/n},\lambda_n \}$ under Theorem 3.2, using the fact that $h_1^2Mk/\alpha_1\le 1/(2\tau)$ and $h_2^2Mk/\alpha_1\le 1/(2\tau)$ respectively. This implies a tightening of the bounds as $M$ increases and directly highlights the benefits of integrative learning.

Although Theorems \ref{thm:MEbound_gbridge}-\ref{thm:MEbound_glasso} establish statistical error bounds for any global optimum of (\ref{eqt:corrected}), it is not immediately clear how to computationally obtain such an optimum. This is partly due to non-convexity in the loss function, which hinders a closed form solution to (\ref{eqt:corrected}). Hence one needs to resort to some type of projected gradient descent algorithm in order to approximate the solution, which generates a sequence of iterates via the following recursions:
\begin{eqnarray}
\label{eqt:pgrad}
{\bm \eta}^{(t+1)} = \argmin_{\rho({\bm\eta})\le R}\bigg\{ \mathcal{L}({\bm \eta}^{(t)} + \langle \hat{\bm{\Gamma}}{\bm \eta} - \hat{\bm \gamma}, {\bm \eta}-{\bm \eta}^{(t)} \rangle) + \frac{\delta^*}{2} \|{\bm \eta}-{\bm \eta}^{(t)} \|^2_2 \bigg\},
\end{eqnarray}
where $\delta^*$ denotes the step size. However for non-convex problems, the projected gradient descent may get trapped in local minima. While some local optima may lie close to the global optimum as per Remark 3.1, not all local optima are guaranteed to converge to optimum solutions that satisfy the statistical error bounds in Theorems \ref{thm:MEbound_gbridge}-\ref{thm:MEbound_glasso}. Fortunately, it is possible to show that the local optima under the projected gradient descent algorithm in (\ref{eqt:pgrad}) involving the $L_{1,q}$ penalty converges (after a suitable number of iterations) to a solution that is arbitrarily close to the global optimum in Theorem \ref{thm:MEbound_glasso}, which ensures the legitimacy of the approximate solution. We first state an additional upper restricted eigen value (upper-RE) condition below that holds with high probability for our settings (Lemma 2 in Supplementary Materials) and is needed in order to derive such a result, followed by the Theorem statement.  We note that it is possible to check whether the upper-RE and lower-RE hold in practice using certain sufficient conditions as described in the Supplementary Materials.
\begin{definition}
(Upper-RE condition). The matrix $\hat{\bm{\Gamma}}$ satisfies an upper restricted eigenvalue condition with curvature $\alpha_2>0$ and tolerance $\tau>0$ if
$ \bm{\theta}^T\hat{\bm{\Gamma}}\bm{\theta}\le \alpha_2\|\bm{\theta}\|_2^2 + \tau\|\bm{\theta}\|_1^2, \forall \bm{\theta}\in\mathbb{R}^{p}.$
\end{definition}



\begin{theorem}(optimization error) \label{thm:optimbound}
Let $\hat{\bm{\eta}}$ denote an optimum solution in Theorem \ref{thm:MEbound_glasso} under $L_{1,q}$ penalty. Then, the estimate $\bm{\eta}^{(t)}$  under the projected gradient descent in (\ref{eqt:pgrad}) with initial choice $\bm{\eta}^*$ satisfies the error bound $\small \|\bm{\eta}^{(t)} - \hat{\bm{\eta}}\|_2^2\le c_3\frac{k\log p}{n}\|\hat{\bm{\eta}}-\bm{\eta}^0\|_2^2$ for all iterates $\small t\ge 2[\log(1/\kappa)]^{-1}\log\frac{\mathcal{L}(\bm{\eta}^*)-\mathcal{L}(\hat{\bm{\eta}})}{\delta^2} + \log_2\log_2(\frac{R\lambda_n}{\delta^2})(1+\frac{\log 2}{\log(1/\kappa)})$ with probability at least $1-c_1\exp\{-c_2\log p \}$, for positive constants $c_1,c_2,c_3>0$, $\delta^2=c_3\frac{k\log p}{n}\|\hat{\bm{\eta}}-\bm{\eta}^0\|_2^2$, and  $\kappa\in(0,1)$.
\end{theorem}


Theorem \ref{thm:optimbound} shows that the $L_2$ optimization error under the projected gradient descent algorithm with $L_{1,q}$ penalty is bounded by the statistical error, which is already shown to be well behaved and bounded in Theorem \ref{thm:MEbound_glasso}. This result essentially guarantees that the iterate $t$ under the projected gradient descent that is easily computed in polynomial-time (especially when $q=2$ resulting in group lasso penalty), converges to a global optimum for the criteria (\ref{eqt:corrected}) that may be difficult to compute and may not have a closed form solution. In other words, Theorem \ref{thm:optimbound} guarantees that when the gradient descent is run long enough, the iterations under the projected gradient descent will produce an estimate that is essentially as good as any global optimum for (\ref{eqt:corrected}) in terms of statistical error. This is indeed a desirable practical feature in our applications of interest. Moreover, the number of iterations needed to be run before the approximate solution starts to converge to the optimal solution will depend on the initial choice of the parameters $\bm{\eta}^*$, with a good choice resulting in faster convergence. This is clear from the fact that a choice of ${\bm \eta^*}$ close to the global optimum will result in a small difference in $\mathcal{L}(\bm{\eta}^*)-\mathcal{L}(\hat{\bm{\eta}})$, subject to the curvature of the loss function. 

\vskip 3pt

{\noindent \emph{Case with unknown noise covariance:}} In practical scenarios, $\Sigma_u$ is unknown and needs to be estimated. Fortunately, under certain scenarios involving replicated validation data, it is possible to empirically estimate the noise covariance in a manner that ensures that the theoretical properties are preserved. In particular, if we observe $n_0$ i.i.d. noise vectors ${\bf u}$, or in the case of repeated observations from healthy controls, i.e. $\bm{z}_{mi} = \bm{x}_i + \bm{u}_{mi}$ with  $\bm{u}_{mi}\stackrel{i.i.d.}{\sim}N(0,\bm{\Sigma}_u)$, Theorems \ref{thm:MEbound_gbridge}-\ref{thm:optimbound} will hold under empirical estimates of ${\bm \Sigma}_u$ provided that the validation data has a reasonably large sample size, as stated below.


\begin{corollary}
Theorems 3.1-3.3 hold if we replace $\bm{\Sigma}_u$ in (\ref{eqt:corrected}) by the estimate $\hat{\bm{\Sigma}}_u=\frac{1}{n_0}U_0^T U_0$ as well as $\small
\hat{\bm{\Sigma}}_u = \frac{1}{n^*(M-1)}\sum_{i=1}^{n^*}\sum_{m=1}^M (\bm{z}_{mi} - \bar{\bm{z}}_{\cdot i})(\bm{z}_{mi} - \bar{\bm{z}}_{\cdot i})^T$, where $\bar{\bm{z}}_{\cdot i} = \sum_{m=1}^M \bm{z}_{mi} /M$, with $\bm{z}_{mi} = \bm{x}_i + \bm{u}_{mi}$ for healthy controls  and $\bm{u}_{mi}\stackrel{i.i.d.}{\sim}N(0,\bm{\Sigma}_u)$ for the second estimator, given that $n_0>n$ and $n^*(M-1)>n$.
\label{coro:unknown}
\end{corollary}

The proof of the above result (provided in the Supplementary Materials) proceeds by showing that the deviation condition in Lemma \ref{thm:deviation} as well as the lower- and upper-RE conditions still hold with high probability under the modified covariance estimator $\hat{\bm{\Sigma}}_u$ in Corollary \ref{coro:unknown}. We use this strategy in our analysis of ADNI data that comprises longitudinal visits for healthy controls in addition to individuals with AD, as detailed in Section 5. 

\vskip 12pt 

{\noindent \bf 3.2. Computational Algorithms}
\label{subsec:noisy_comp}

We implement the projected gradient descent approach under the group bridge and the group lasso penalties, whose convergence  can be impacted by the choice of the step size $\delta^*$ in (\ref{eqt:pgrad}). We utilize the non-monotone spectral projected gradient (SPG) method for the computation under the group lasso penalty as in \cite{sra2011fast} and \cite{duchi2008efficient} (see Supplementary Materials for details), which adopts a spectral choice of the step size with non-monotone line search technique that is known to speed up the convergence. Moreover under the non-convex group bridge penalty, we develop a novel projected gradient descent algorithm which restricts the space of admissible solutions to an $L_1$ ball of radius $R^2$ by leveraging Corollary \ref{thm:coro_L1restriction}.  Using the data augmentation strategy in  \cite{huang2009group}, criteria (\ref{eqt:corrected}) under group bridge can be expressed as the following equivalent problem :
\begin{equation}
    \underset{\|\bm{\eta}\|_1\le R^2}{\min} \bigg[\sum_{m=1}^M\bigg\{\frac{1}{2}\bm{\eta}_m^T\hat{\bm{\Gamma}}_m\bm{\eta}_m - \langle\hat{\bm{\gamma}}_m,\bm{\eta}_m\rangle\bigg\} + \sum_{j=1}^p\theta_j^{-1}\Big(\sum_{m=1}^M|\eta_{mj}|\Big) + \tau_n\sum_{j=1}^p\theta_j\bigg], \mbox{ } \tau_n = \lambda_n^2/4,
\label{eqt:corrected_bridge}
\end{equation}
where $\theta_j\ge 0$ for $j=1,\cdots,p$. This can be solved via the SPG method. Details can be found in Algorithm 5 of the Supplementary Materials.

In practice, the radius $R$ in (\ref{eqt:corrected}) and  (\ref{eqt:corrected_bridge}) can be set to be relatively large to ensure a reasonable bound of the estimate. One can get a better sense about the lower bound of the constraints $R$ and $R^2$ from the estimates of the group lasso and group bridge methods without noise correction in (\ref{eqt:max4}),  which can be used to guide the choice of $R$. More discussions on the choice of $R$ can be found in the Supplementary Materials. Alternatively, the correction proposed in \cite{datta2017cocolasso} avoids tuning on $R$ and the projection step, which may be potentially interesting to consider in future work. Moreover, the shrinkage parameter $\lambda_n$ in the penalty term can be selected via five fold cross validation. In our implementations, we fix the primary level of wavelet transform ($j_0$) informed by extensive empirical studies. 
However, in the situation when one is uncertain about the choice of $j_0$, a cross validation can be conducted based on cross-validation or goodness of fit scores.




\section{Simulations}
\label{sec:sims}
\begin{figure}
    \centering
    \includegraphics[width=\textwidth]{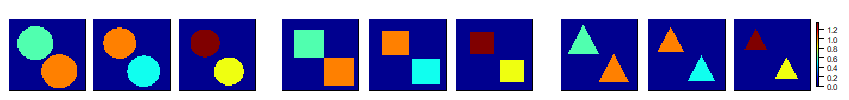}
    \caption{Partially overlapping true signals used in simulations, where the size of the signals varying across data sources. Other signals types (homogeneous and minimally overlapping) were also considered with these results presented in the Supplementary Materials.}
    \label{fig:truesignal}
\end{figure}

In this section we conduct extensive simulations involving three data sources ($M=3$) with 2-D images of size $64\times 64$ that mimics the 2-D brain slices and evaluate the performance of the proposed approach with respect to competing methods. The functional image predictors are generated by first generating wavelet coefficients independently from normal distribution with mean 0 and variance 1, and followed by an inverse wavelet transform implemented via the \texttt{R} package \texttt{wavethresh} \citep{nason2008wavelet}. We choose the Daubechies Least Asymmetric wavelet with 4 vanishing moments and $j_0=3$ as the wavelet basis function in both data generation and model fitting under all wavelet-based approaches. We generated three types of true 2-D functional regression coefficients with different shapes including round, square and triangle, and varying degrees of overlap in the regions with non-zero signals between the three images across the data sources as shown in Figure \ref{fig:truesignal}. We also considered two extreme simulation settings with homogeneous and minimally-overlapping signals, but we present these results in the Supplementary Materials due to space constraints. The scalar outcome variable is then generated under a scalar-on-image regression model based on the true image without noise. The ratio of the mean function variance and residual term variance is set to be 9. For model fitting, we use a working image that is obtained by corrupting the true image with additive noise, and explore the performance for both known and unknown error covariances. We simulate data with training and test sample sizes as 200 for all three groups.  

{\noindent \emph{Competing Methods and Evaluation Metrics:}}
 We compare the performance of the proposed projected group lasso (‘p\_glasso’) and the projected group bridge (‘p\_gbridge’) methods with several competing methods that (i) fit the model separately for each data source without accounting for noise in images, including 
the WNET method \citep{reiss2015wavelet} which first performs wavelet transform and then applies elastic net penalty regression, and the WPCR method \citep{reiss2015wavelet} which first performs wavelet transform then applies principal component regression; (ii) fit the model separately to each data source with noise correction, as in the method in \cite{loh2012} that uses a similar projected gradient descent algorithm as in (\ref{eqt:pgrad}), but with $L_1$ penalty on the wavelet coefficients that is not equipped for multi-task learning (`p\_lasso'); and (iii) multi-task learning approaches involving group lasso and group bridge penalties, but without noise correction as in (\ref{eqt:max4}). The 
WNET and WPCR methods are implemented in \texttt{R} package \texttt{refund.wave} \citep{reiss2015wavelet}.  We utilize the \texttt{R} package \texttt{grpreg} \citep{grpreg} for implementing the group lasso without noise correction, while we use a hierarchical representation to implement the group bridge without measurement error, as detailed in the Supplementary Materials. We evaluate the performance of different methods using  out-of-sample prediction in a testing sample measured via the prediction mean squared error (PMSE), as well as the accuracy in recovering the functional regression coefficient in terms of bias and area under the curve (AUC) that measures the ability to distinguish non-zero and zero signals (results in Table \ref{tab:sim2d}). The PMSE is calculated for the test set and standardized by the variance of the training set. Figures 1-3 in Supplementary Materials illustrate the recovery of the true signals for $SNR=0$ and $SNR=3$, while Table \ref{tab:sim2d} presents replicate-averaged PMSE, AUC and bias.

\vskip 3pt

{\noindent \bf \large 4.1. Scenario with Known Noise Covariance}

 The true images are generated independently for each of the data sources, and additive noise is  introduced to the true image to obtain the working image used for model fitting. The noise is generated by first generating wavelet coefficients independently from normal distribution with mean 0 and variance at 1/4 (SNR=4) or 1/3 (SNR= 3), and then followed by an inverse wavelet transform. Results are also reported for an ideal case where the true image is used to fit the data. We report results averaged over 100 replicates for each setting.


{\noindent \emph{Results:}} For the scenarios where the true images are observed, the group bridge method without noise correction shows the best predictive performance. However, the group lasso approach without noise correction, as well as the projected group lasso methods with noise correction have comparable performance with respect to the group bridge method, in terms of coefficient estimation (bias and AUC) that is superior to remaining methods. While it is expected that methods without noise correction will have improved predictive performance when in fact the true image is observed, it is impressive to see that the projected group lasso approach with noise correction performs equally well in terms of signal recovery under these settings, although it can't guarantee accurate prediction due to the assumption of noise in the images when there is in fact none. For settings of greater interest involving noisy images used for model fitting, the projected group lasso approach  has (by far) the best performance in terms of out of sample prediction, bias and AUC, which are almost always significantly improved compared to competing methods. While the empirical performance of the projected group lasso approach in terms of coefficient estimation is supported by our theoretical results, the superior predictive performance under the projected group lasso approach in the presence of noisy images further highlights the utility of this method. In contrast, the signal recovery accuracy under the projected group bridge is variable and the prediction performance is not always ideal, which is likely due to the lack of theoretical guarantees under the projected gradient descent algorithm that used to compute parameter estimates corresponding to the group bridge penalty. Figures 1-3 in Supplementary Materials illustrate that while the proposed approaches are able to adequately recover the true signals, some competing approaches (such as WPCR) have a particularly poor signal recovery. 

\vskip 5pt


{\noindent \bf \large 4.2. Scenario with Unknown Noise Covariance}

In this scenario, 
we assume $\Sigma_u$ is unknown and needs to be estimated from  an external validation sample, which is assumed to be a valid estimate for the noise covariance in the training and the test data sets as well, as in our motivating ADNI analysis. This set-up is designed to mimic the scenario for ADNI data analysis (see Section 5 for more details), and does not use the validation data to inform any other aspect of the modeling or prediction conducted on the training and test data sets beyond computing the error covariance.   Unlike the set-up with known noise-covariance, the images for each sample were linked across the three data sources by first generating a true image independently for each sample, and then corrupting these images with additive noise across the three data sources. This scenario leads to common patterns in images across the three data sources that are distorted by noise, and enables one to empirically estimate the noise covariance. We generate the data with varying signal-to-noise ratios (6, 4, or 3) and for 100 replicates for each simulation set-up.  



{\noindent \emph{Results:}} The projected group lasso method consistently has the best prediction performance across all signal shapes and signal-to-noise ratios that is significantly improved compared to the other approaches. In terms of coefficient estimation, both the projected group bridge as well as the projected group lasso methods consistently have significantly improved performance in terms of bias and AUC compared to the other methods. Compared to the case with known noise covariance, the relative performance (bias and AUC) under the projected group lasso method slightly deteriorates in the case of unknown noise covariance and becomes more at par with the projected group bridge in terms of signal recovery. However in our experience, the performance of the  projected group bridge was occasionally sensitive to the starting values, whereas the projected group lasso method produces more stable results that is consistent with the theoretical guarantees under Theorem \ref{thm:optimbound}. 



{\noindent {\em Summary of Results}:} The results clearly illustrate the benefits of multi-task learning in scalar-on-image regression, given that the performance of those approaches that fit the model separately for each data source are often inferior, even when they account for the presence of noise. In addition, multi-task learning without noise correction results in sub-optimal performance in the presence of noisy images, compared to projected group lasso and group bridge approaches. Moreover, the advantages of noise correction under grouped penalties are accentuated under homogeneous signals, and are partially eroded under minimally overlapping true signals across data sources, as expected (see Supplementary Material). However, the proposed projected group lasso and/or group bridge still have improved or comparable predictive and estimation performance across the overwhelming majority of settings for the latter case. 
  Another factor that influences the performance of the proposed methods is the accuracy of the estimated ${\bm \Sigma}_u$ used in the working model. In our experience, the prediction and selection performance are largely robust to mis-specification of the noise covariance, as long as the biases for the estimated noise covariance are not overly pronounced (see additional simulations in Supplementary Materials). In practical applications, the proposed approach is best suited in settings where validation datasets with non-negligible sample sizes are available for estimating the unknown ${\bm \Sigma}_u$. For our simulations, the proposed methods with noise correction often converge within 3 minutes on a machine with 1.90GHz Intel i7 processor and 16GB RAM. Figure 4 in Supplementary Materials shows the convergence plots. 


\begin{table}[htbp!]
\centering
\resizebox{0.925\textwidth}{!}{
\begin{tabular}{|c|c|l||ccc|ccc|ccc||ccc|ccc|ccc||}
\hline
\multicolumn{3}{|c||}{} & \multicolumn{9}{|c||}{\textit{Known Noise Covariance}} & \multicolumn{9}{|c||}{\textit{Unknown Noise Covariance}} \\
\cline{4-21}
\multicolumn{3}{|c||}{} & \multicolumn{3}{c|}{Noiseless} & \multicolumn{3}{c|}{SNR=4} & \multicolumn{3}{c||}{SNR=3} & \multicolumn{3}{c|}{SNR=6} & \multicolumn{3}{c|}{SNR=4} & \multicolumn{3}{c||}{SNR=3} \\
\cline{4-21}
\multicolumn{3}{|c||}{} & G1 & G2 & G3 & G1 & G2 & G3 & G1 & G2 & G3 & G1 & G2 & G3 & G1 & G2 & G3 & G1 & G2 & G3 \\
\hline
\hline
\multirow{21}{*}{\rotatebox[origin=c]{90}{PMSE}} & \multirow{7}{*}{\rotatebox[origin=c]{90}{round}}
& WNET & 0.67 & 0.58 & 0.51 & 0.83 & 0.77 & 0.75 & 0.89 & 0.82 & 0.77 & 0.81 & 0.73 & 0.69 & 0.85 & 0.77 & 0.72 & 0.89 & 0.82 & 0.78 \\
& & WPCR & 1.01 & 1.01 & 1.01 & 1.01 & 1.01 & 1.01 & 1.01 & 1.01 & 1.01 & 1.01 & 1.01 & 1.01 & 1.01 & 1.01 & 1.01 & 1.01 & 1.01 & 1.01 \\
& & p\_lasso & 0.68 & 0.60 & 0.53 & 0.81 & 0.74 & 0.70 & 0.92 & 0.83 & 0.74 & 0.77 & 0.68 & 0.62 & 0.83 & 0.72 & 0.67 & 0.89 & 0.79 & 0.74 \\
& & glasso & 0.42 & 0.36 & 0.33 & 0.65 & 0.57 & 0.56 & 0.71 & 0.63 & 0.61 & 0.73 & 0.66 & 0.62 & 0.80 & 0.71 & 0.68 & 0.84 & 0.77 & 0.74 \\
& & gbridge & \bf{0.32} & \bf{0.34} & \bf{0.29} & 0.71 & 0.67 & 0.64 & 0.81 & 0.74 & 0.69 & \bf{0.64} & 0.63 & 0.56 & 0.77 & 0.72 & 0.68 & 0.87 & 0.79 & 0.74 \\
& & p\_glasso & 0.44 & 0.37 & 0.35 & \bf{0.58} & \bf{0.52} & \bf{0.50} & \bf{0.67} & \bf{0.60} & \bf{0.58} & \bf{0.62} & \bf{0.54} & \bf{0.49} & \bf{0.69} & \bf{0.57} & \bf{0.55} & \bf{0.75} & \bf{0.63} & \bf{0.61} \\
& & p\_gbridge & 0.47 & 0.43 & 0.37 & 0.72 & 0.65 & 0.65 & 0.83 & 0.73 & 1.00 & 0.70 & 0.59 & 0.55 & 0.81 & 0.61 & 0.63 & 0.89 & 0.70 & 0.71 \\
\cline{2-21}
& \multirow{7}{*}{\rotatebox[origin=c]{90}{square}}
& WNET & 0.69 & 0.59 & 0.53 & 0.84 & 0.77 & 0.74 & 0.88 & 0.81 & 0.78 & 0.79 & 0.73 & 0.69 & 0.84 & 0.78 & 0.72 & 0.86 & 0.82 & 0.78 \\
& & WPCR & 1.01 & 1.01 & 1.01 & 1.01 & 1.01 & 1.01 & 1.01 & 1.01 & 1.01 & 1.01 & 1.01 & 1.01 & 1.01 & 1.01 & 1.01 & 1.01 & 1.01 & 1.01 \\
& & p\_lasso & 0.69 & 0.61 & 0.54 & 0.81 & 0.73 & 0.70 & 0.88 & 0.79 & 0.75 & 0.75 & 0.68 & 0.63 & 0.81 & 0.73 & 0.67 & 0.84 & 0.79 & 0.77 \\
& & glasso & 0.45 & 0.37 & 0.35 & 0.67 & 0.59 & 0.57 & 0.71 & 0.63 & \bf{0.61} & 0.72 & 0.65 & 0.62 & 0.79 & 0.72 & 0.68 & 0.83 & 0.78 & 0.75 \\
& & gbridge & \bf{0.36} & \bf{0.35} & \bf{0.32} & 0.73 & 0.67 & 0.65 & 0.80 & 0.73 & 0.71 & \bf{0.63} & 0.60 & 0.58 & 0.74 & 0.71 & 0.66 & 0.84 & 0.78 & 0.75 \\
& & p\_glasso & 0.47 & 0.39 & 0.36 & \bf{0.63} & \bf{0.52} & \bf{0.52} & \bf{0.68} & \bf{0.59} & \bf{0.60} & \bf{0.61} & \bf{0.53} & \bf{0.50} & \bf{0.68} & \bf{0.59} & \bf{0.55} & \bf{0.72} & \bf{0.64} & \bf{0.64} \\
& & p\_gbridge & 0.52 & 0.44 & 0.40 & 1.13 & 0.61 & 0.66 & 0.86 & 0.78 & 0.81 & 0.72 & 0.56 & 0.56 & 0.80 & 0.64 & 0.64 & 0.83 & 0.70 & 0.71 \\
\cline{2-21}
& \multirow{7}{*}{\rotatebox[origin=c]{90}{triangle}}
& WNET & 0.46 & 0.54 & 0.55 & 0.67 & 0.72 & 0.70 & 0.72 & 0.75 & 0.74 & 0.60 & 0.66 & 0.65 & 0.66 & 0.71 & 0.70 & 0.71 & 0.75 & 0.74 \\
& & WPCR & 1.01 & 1.01 & 1.01 & 1.01 & 1.02 & 1.01 & 1.01 & 1.01 & 1.01 & 1.01 & 1.01 & 1.00 & 1.01 & 1.01 & 1.01 & 1.01 & 1.01 & 1.01 \\
& & p\_lasso & 0.48 & 0.56 & 0.55 & 0.62 & 0.70 & 0.66 & 0.68 & 0.73 & 0.71 & 0.55 & 0.61 & 0.61 & 0.61 & 0.66 & 0.66 & 0.67 & 0.70 & 0.71 \\
& & glasso & 0.36 & 0.39 & 0.41 & 0.57 & 0.58 & 0.59 & 0.62 & 0.63 & 0.62 & 0.56 & 0.60 & 0.58 & 0.64 & 0.66 & 0.65 & 0.69 & 0.71 & 0.69 \\
& & gbridge & \bf{0.30} & \bf{0.35} & \bf{0.39} & 0.55 & 0.58 & 0.59 & 0.59 & 0.62 & 0.63 & 0.48 & 0.54 & 0.53 & 0.56 & 0.60 & 0.63 & 0.62 & 0.65 & 0.67 \\
& & p\_glasso & 0.38 & 0.40 & 0.42 & \bf{0.51} & \bf{0.55} & \bf{0.56} & \bf{0.57} & \bf{0.59} & \bf{0.61} & \bf{0.46} & \bf{0.51} & \bf{0.51} & \bf{0.52} & \bf{0.56} & \bf{0.57} & \bf{0.58} & \bf{0.61} & \bf{0.63} \\
& & p\_gbridge & 0.42 & 0.49 & 0.50 & 0.59 & 0.62 & 0.66 & 0.69 & 0.68 & 0.72 & 0.55 & 0.60 & 0.64 & 0.58 & 0.60 & 0.67 & 0.63 & 0.66 & 0.70 \\
\hline
\hline
\multirow{21}{*}{\rotatebox[origin=c]{90}{Bias}} & \multirow{7}{*}{\rotatebox[origin=c]{90}{round}}
& WNET & 0.25 & 0.19 & 0.21 & 0.27 & 0.20 & 0.24 & 0.27 & 0.21 & 0.25 & 0.27 & 0.20 & 0.23 & 0.27 & 0.21 & 0.24 & 0.27 & 0.21 & 0.24 \\
& & WPCR & 0.29 & 0.24 & 0.27 & 0.29 & 0.23 & 0.27 & 0.29 & 0.23 & 0.27 & 0.29 & 0.23 & 0.27 & 0.29 & 0.23 & 0.27 & 0.29 & 0.23 & 0.27 \\
& & p\_lasso & 0.25 & 0.19 & 0.21 & 0.25 & 0.19 & 0.22 & 0.26 & 0.20 & 0.22 & 0.25 & 0.19 & 0.21 & 0.25 & 0.19 & 0.22 & 0.26 & 0.20 & 0.22 \\
& & glasso & 0.18 & \bf{0.13} & \bf{0.16} & 0.23 & 0.17 & 0.21 & 0.24 & 0.18 & 0.22 & 0.26 & 0.20 & 0.24 & 0.27 & 0.21 & 0.24 & 0.28 & 0.22 & 0.26 \\
& & gbridge & \bf{0.17} & 0.14 & 0.17 & 0.23 & 0.19 & 0.23 & 0.24 & 0.19 & 0.23 & 0.23 & 0.19 & 0.23 & 0.24 & 0.20 & 0.24 & 0.26 & 0.21 & 0.24 \\
& & p\_glasso & 0.18 & 0.14 & 0.17 & \bf{0.19} & \bf{0.14} & \bf{0.17} & \bf{0.20} & \bf{0.14} & \bf{0.18} & {\bf 0.21} & 0.16 & 0.19 & \bf{0.21} & {\bf 0.15} & {\bf 0.18} & \bf{0.22} & {\bf 0.16} & 0.19 \\
& & p\_gbridge & 0.20 & 0.16 & 0.19 & 0.19 & 0.15 & 0.18 & 0.21 & 0.16 & 0.20 & \bf{0.21} & \bf{0.15} & \bf{0.18} & \bf{0.21} & \bf{0.15} & \bf{0.18} & \bf{0.22} & \bf{0.16} & \bf{0.18} \\
\cline{2-21}
& \multirow{7}{*}{\rotatebox[origin=c]{90}{square}}
& WNET & 0.22 & 0.16 & 0.18 & 0.23 & 0.18 & 0.20 & 0.24 & 0.18 & 0.20 & 0.23 & 0.17 & 0.19 & 0.24 & 0.18 & 0.20 & 0.24 & 0.18 & 0.20 \\
& & WPCR & 0.26 & 0.20 & 0.23 & 0.25 & 0.19 & 0.23 & 0.25 & 0.19 & 0.22 & 0.26 & 0.19 & 0.22 & 0.26 & 0.19 & 0.23 & 0.25 & 0.19 & 0.22 \\
& & p\_lasso & 0.22 & 0.16 & 0.18 & 0.22 & 0.16 & 0.19 & 0.23 & 0.17 & 0.18 & 0.22 & 0.16 & 0.18 & 0.22 & 0.16 & 0.18 & 0.22 & 0.17 & 0.19 \\
& & glasso & \bf{0.17} & \bf{0.12} & \bf{0.15} & 0.20 & 0.15 & 0.18 & 0.21 & 0.15 & 0.18 & 0.23 & 0.17 & 0.20 & 0.24 & 0.18 & 0.21 & 0.25 & 0.19 & 0.22 \\
& & gbridge & \bf{0.17} & 0.13 & 0.17 & 0.22 & 0.17 & 0.21 & 0.22 & 0.17 & 0.20 & 0.21 & 0.17 & 0.21 & 0.22 & 0.17 & 0.21 & 0.23 & 0.18 & 0.21 \\
& & p\_glasso & {\bf 0.17} & {\bf 0.12} & {\bf 0.15} & \bf{0.17} & \bf{0.12} & \bf{0.15} & \bf{0.18} & \bf{0.13} & \bf{0.15} & \bf{0.19} & 0.14 & \bf{0.16} & \bf{0.19} & 0.14 & \bf{0.16} & \bf{0.19} & 0.14 & \bf{0.17} \\
& & p\_gbridge & 0.19 & 0.15 & 0.17 & 0.19 & 0.13 & 0.16 & 0.19 & 0.14 & 0.16 & \bf{0.19} & \bf{0.13} & \bf{0.16} & \bf{0.19} & \bf{0.13} & \bf{0.16} & \bf{0.19} & \bf{0.13} & \bf{0.16} \\
\cline{2-21}
& \multirow{7}{*}{\rotatebox[origin=c]{90}{triangle}}
& WNET & 0.11 & 0.09 & 0.10 & 0.12 & 0.10 & 0.10 & 0.12 & 0.09 & 0.10 & 0.12 & 0.09 & 0.10 & 0.12 & 0.10 & 0.10 & 0.12 & 0.10 & 0.10 \\
& & WPCR & 0.14 & 0.10 & 0.11 & 0.14 & 0.11 & 0.11 & 0.14 & 0.10 & 0.11 & 0.14 & 0.11 & 0.11 & 0.14 & 0.11 & 0.11 & 0.14 & 0.10 & 0.11 \\
& & p\_lasso & 0.11 & 0.09 & 0.10 & 0.11 & 0.09 & 0.09 & 0.11 & 0.09 & 0.09 & 0.11 & 0.09 & 0.10 & 0.11 & 0.09 & 0.09 & 0.11 & 0.09 & \bf{0.09} \\
& & glasso & 0.09 & \bf{0.08} & \bf{0.09} & 0.11 & 0.08 & 0.09 & 0.11 & 0.09 & 0.09 & 0.12 & 0.09 & 0.10 & 0.13 & 0.10 & 0.11 & 0.13 & 0.10 & 0.11 \\
& & gbridge & \bf{0.09} & 0.08 & 0.10 & 0.11 & 0.08 & 0.10 & 0.11 & 0.08 & 0.10 & 0.11 & 0.09 & 0.11 & 0.11 & 0.09 & 0.11 & 0.12 & 0.09 & 0.11 \\
& & p\_glasso & 0.10 & \bf{0.08} & \bf{0.09} & \bf{0.09} & \bf{0.08} & \bf{0.09} & \bf{0.10} & \bf{0.08} & \bf{0.09} & \bf{0.10} & \bf{0.08} & \bf{0.09} & \bf{0.10} & \bf{0.08} & \bf{0.09} & \bf{0.10} & \bf{0.08} & \bf{0.09} \\
& & p\_gbridge & 0.12 & 0.10 & 0.12 & 0.10 & 0.08 & 0.10 & \bf{0.10} & \bf{0.08} & 0.09 & 0.11 & 0.09 & 0.11 & \bf{0.10} & \bf{0.08} & 0.10 & \bf{0.10} & \bf{0.08} & \bf{0.09} \\
\hline
\hline
\multirow{21}{*}{\rotatebox[origin=c]{90}{AUC}} & \multirow{7}{*}{\rotatebox[origin=c]{90}{round}}
& WNET & 0.81 & 0.81 & 0.89 & 0.76 & 0.77 & 0.83 & 0.74 & 0.76 & 0.82 & 0.77 & 0.78 & 0.85 & 0.76 & 0.76 & 0.83 & 0.73 & 0.75 & 0.83\\
& & WPCR & 0.53 & 0.53 & 0.54 & 0.62 & 0.62 & 0.61 & 0.62 & 0.61 & 0.61 & 0.62 & 0.61 & 0.60 & 0.62 & 0.62 & 0.61 & 0.63 & 0.62 & 0.61\\
& & p\_lasso & 0.79 & 0.80 & 0.89 & 0.77 & 0.78 & 0.85 & 0.74 & 0.73 & 0.84 & 0.79 & 0.80 & 0.87 & 0.78 & 0.77 & 0.84 & 0.74 & 0.75 & 0.83\\
& & glasso & 0.94 & \bf{0.95} & \bf{0.97} & 0.90 & 0.92 & 0.93 & 0.88 & 0.90 & 0.92 & 0.81 & 0.84 & 0.87 & 0.79 & 0.81 & 0.85 & 0.76 & 0.78 & 0.83\\
& & gbridge & \bf{0.95} & 0.94 & \bf{0.97} & 0.86 & 0.86 & 0.90 & 0.85 & 0.85 & 0.88 & 0.85 & 0.86 & 0.91 & 0.84 & 0.84 & 0.88 & 0.82 & 0.82 & 0.87\\
& & p\_glasso & 0.94 & \bf{0.95} & \bf{0.97} & \bf{0.93} & \bf{0.94} & \bf{0.96} & \bf{0.92} & \bf{0.93} & \bf{0.95} & 0.89 & 0.90 & 0.93 & 0.90 & \bf{0.91} & \bf{0.93} & \bf{0.89} & \bf{0.89} & \bf{0.92}\\
& & p\_gbridge & 0.90 & 0.91 & 0.94 & 0.91 & 0.91 & 0.94 & 0.88 & 0.87 & 0.90 & \bf{0.90} & \bf{0.91} & \bf{0.94} & \bf{0.91} & \bf{0.91} & \bf{0.93} & \bf{0.89} & \bf{0.89} & \bf{0.92}\\
\cline{2-21}
& \multirow{7}{*}{\rotatebox[origin=c]{90}{square}}
& WNET & 0.81 & 0.83 & 0.90 & 0.77 & 0.78 & 0.84 & 0.74 & 0.77 & 0.85 & 0.78 & 0.80 & 0.87 & 0.77 & 0.78 & 0.86 & 0.76 & 0.78 & 0.84\\
& & WPCR & 0.53 & 0.54 & 0.54 & 0.66 & 0.64 & 0.63 & 0.64 & 0.64 & 0.63 & 0.65 & 0.64 & 0.64 & 0.65 & 0.64 & 0.63 & 0.65 & 0.64 & 0.63\\
& & p\_lasso & 0.81 & 0.82 & 0.89 & 0.78 & 0.79 & 0.85 & 0.73 & 0.77 & 0.85 & 0.80 & 0.81 & 0.88 & 0.79 & 0.79 & 0.87 & 0.77 & 0.77 & 0.85\\
& & glasso & \bf{0.94} & \bf{0.95} & \bf{0.97} & 0.90 & 0.91 & 0.94 & 0.89 & 0.91 & 0.93 & 0.82 & 0.85 & 0.89 & 0.79 & 0.81 & 0.86 & 0.77 & 0.79 & 0.84\\
& & gbridge & \bf{0.94} & \bf{0.95} & \bf{0.97} & 0.86 & 0.88 & 0.91 & 0.85 & 0.86 & 0.90 & 0.87 & 0.88 & 0.92 & 0.85 & 0.86 & 0.90 & 0.83 & 0.85 & 0.88\\
& & p\_glasso & \bf{0.94} & \bf{0.95} & \bf{0.97} & \bf{0.93} & \bf{0.94} & \bf{0.95} & \bf{0.92} & \bf{0.93} & \bf{0.95} & 0.90 & 0.91 & \bf{0.94} & \bf{0.90} & \bf{0.90} & \bf{0.93} & \bf{0.90} & \bf{0.90} & \bf{0.92}\\
& & p\_gbridge & 0.89 & 0.91 & 0.94 & 0.91 & 0.91 & 0.93 & 0.89 & 0.88 & 0.91 & \bf{0.91} & \bf{0.92} & \bf{0.94} & \bf{0.90} & \bf{0.91} & \bf{0.94} & \bf{0.90} & \bf{0.90} & \bf{0.93}\\
\cline{2-21}
& \multirow{7}{*}{\rotatebox[origin=c]{90}{triangle}}
& WNET & 0.92 & 0.90 & 0.94 & 0.89 & 0.86 & 0.92 & 0.87 & 0.85 & 0.91 & 0.90 & 0.89 & 0.93 & 0.88 & 0.86 & 0.91 & 0.87 & 0.85 & 0.91\\
& & WPCR & 0.54 & 0.55 & 0.56 & 0.66 & 0.65 & 0.65 & 0.66 & 0.65 & 0.65 & 0.65 & 0.64 & 0.64 & 0.66 & 0.64 & 0.65 & 0.66 & 0.65 & 0.65\\
& & p\_lasso & 0.92 & 0.90 & 0.94 & 0.90 & 0.87 & 0.93 & 0.87 & 0.84 & 0.91 & 0.91 & 0.90 & 0.94 & 0.89 & 0.88 & 0.92 & 0.87 & 0.87 & 0.92\\
& & glasso & {\bf 0.98} & \bf{0.98} & \bf{0.98} & 0.95 & 0.96 & 0.97 & 0.95 & 0.96 & 0.97 & 0.93 & 0.94 & 0.95 & 0.91 & 0.91 & 0.94 & 0.89 & 0.90 & 0.93\\
& & gbridge & \bf{0.98} & \bf{0.98} & \bf{0.98} & 0.95 & 0.95 & 0.97 & 0.95 & 0.95 & 0.96 & \bf{0.96} & 0.96 & 0.97 & 0.94 & 0.95 & 0.96 & 0.94 & 0.94 & 0.95\\
& & p\_glasso & \bf{0.98} & \bf{0.98} & \bf{0.98} & \bf{0.97} & \bf{0.97} & \bf{0.98} & \bf{0.96} & \bf{0.97} & \bf{0.97} & \bf{0.96} & \bf{0.96} & \bf{0.97} & \bf{0.96} & \bf{0.96} & \bf{0.97} & \bf{0.95} & \bf{0.96} & \bf{0.97}\\
& & p\_gbridge & 0.95 & 0.95 & 0.96 & 0.95 & 0.95 & 0.96 & 0.95 & 0.95 & 0.97 & \bf{0.96} & 0.96 & 0.96 & \bf{0.95} & \bf{0.95} & 0.96 & \bf{0.95} & 0.95 & 0.96\\
\hline
\end{tabular}}
\caption{Summary for simulation results with known and unknown noise covariances}
\label{tab:sim2d}
\end{table}


\section{Analysis of ADNI Data}
\label{sec:adni}
The Alzheimer’s Disease Neuroimaging Initiative (ADNI) longitudinal study is designed to develop and validate neuroimaging, clinical and genetic biomarkers in clinical trials of Alzheimer’s disease (AD) therapies \citep{weiner2015introduction}.
The primary goal of the ADNI analysis in this article is to discover neuroimaging biomarkers in the form of localized brain regions that are significantly related to longitudinal changes in cognition for AD individuals, using magnetic resonance imaging (MRI) scans that measure the brain structure and brain volumes at the voxel level with dimensions $256\times256\times 170$. 

{\noindent \emph{Data Pre-processing}:}
Our analysis used 1.5T T1-weighted MRI volumetric scans from ADNI-1, created by the ADNI MRI Core. The downloaded data included  MRI scans acquired from 192 healthy controls (NC), and 133 Alzheimer's disease (AD) individuals from screening visit (baseline), month 6 visit and month 12 visit, in addition to age, gender, and APOE status. We pre-processed the MRI scans using the registration pipeline of the Advanced Normalization Tools (ANTs) \citep{avants2011reproducible} and the images were standardized into the space of an ADNI-specific template \citep{tustison2019longitudinal} that addresses the  intra-subject longitudinal variations. More complete details about the pre-processing pipeline for the MRI scans, along with demographic details can be found in Supplementary Materials. 

{\noindent \emph{Analysis Outline:}}
The outcome used for our analysis is the Mini-Mental State Exam (MMSE) score that measures cognitive abilities. We conducted our analysis separately for 9 two-dimensional axial slices each of size $128\times 128$, which covers the hippocampus and amygdala and is our targeted area of interest (depicted in Figure 5 in Supplementary Materials). 
 A supplementary 3-D analysis was also conducted with results in the Supplementary Materials.
Our goal is to study how the relationship between MMSE and brain structures at the voxel level change across time, by jointly analyzing the imaging data across the three longitudinal visits. Due to the fact that age, gender and APOE status did not produce significant associations with the outcome after accounting for the variability due to the brain image, and in order to boost the power to detect important regions, we chose not to adjust for these additional variables in our final scalar-on-image regression model as in \cite{wang2014regularized}. The goals of our analysis are to identify brain regions significantly associated with MMSE, and evaluate the out of sample prediction in the presence of noisy MRI scans under the proposed approaches and the same set of competing approaches as in the simulation studies.

For our analysis, we focus our modeling efforts on 133 AD individuals who have data at baseline, 6 months and 12 months. In addition, we used MRI scans from 192 healthy NC individuals over three longitudinal visits to obtain an estimate for the noise covariance matrix as $\small
\hat{\bm{\Sigma}}_u = \frac{1}{n^*(M-1)}\sum_{i=1}^{n^*}\sum_{m=1}^M (\bm{z}_{mi} - \bar{\bm{z}}_{\cdot i})(\bm{z}_{mi} - \bar{\bm{z}}_{\cdot i})^T$ as in Corollary 3.2, which was subsequently used for the analysis of the AD cohort. 
The extrapolation of the noise covariance from the NC cohort to the AD cohort is valid under the assumption that the noise in the MRI scans is related to scanner properties and does not depend on the disease status or other individual-specific characteristics. We note that the data on the NC individuals was not used to inform the analysis under other competing approaches, since the model parameters under these methods are specific to the analysis of AD individuals and can not be generalized to other cohorts. For model fitting and out of sample prediction under all the methods, we randomly split the 133 AD individuals into training and test groups (50-50), and consider multiple (25) such splits. The significant voxel-level associations were inferred via a two-sided t-test ($\alpha=0.05$) with Bonferroni corrections using the estimated signals over the 25 splits. In order to eliminate clinically weak signals from the association map, all signals with absolute values less than $10^{-3}$ were thresholded to zero before performing the t-test.

\begin{table}[hbt!]
\centering
\resizebox{0.99\textwidth}{!}{
\begin{tabular}{|c|l|ccccccccc|ccccccccc|c|}
\hline
\multicolumn{2}{|c}{ } &\multicolumn{9}{|c}{Prediction PMSE}  &\multicolumn{10}{|c|}{Number of significantly associated voxels} \\
\cline{3-21}
\multicolumn{2}{|c|}{Methods}  & s131 & s130 & s129 & s128 & s127 & s126 & s125 & s124 & s123 & s131 & s130 & s129 & s128 & s127 & s126 & s125 & s124 & s123 & total\\
\hline
\multirow{7}{*}{\rotatebox[origin=c]{90}{Baseline}} 
& WNET & 1.00 & 1.01 & 0.97 & 1.00 & 1.00 & 1.00 & 1.03 & 1.00 & 1.03 &0 & 0 & 0 & 0 & 0 & 0 & 0 & 0 & 7 & 7\\

& WPCR & 1.03 & 1.33 & 1.07 & 1.12 & 1.17 & 1.20 & 1.19 & 1.22 & 1.17 &19 & 0 & 0 & 0 & 0 & 0 & 0 & 56 & 458 & 533\\

& p\_lasso & 0.98 & 1.00 & 1.00 & 1.00 & 1.00 & 1.00 & 1.00 & 1.00 & 1.00 &46 & 0 & 0 & 0 & 0 & 0 & 0 & 0 & 0 & 46\\

& glasso & 0.97 & 0.93 & 0.96 & 0.91 & 0.97 & \bf{0.95} & 0.98 & 0.98 & 0.99 &16 & 0 & 0 & 25 & 0 & 0 & 0 & 0 & 0 & 41\\

& gbridge & 1.00 & 1.00 & 1.00 & 1.00 & 1.00 & 1.00 & 1.00 & 1.00 & 1.00 &0 &0 &0 &0 &0 &0 &0 &0 &0 &0\\

& p\_glasso & \bf{0.92} & \bf{0.86} & \bf{0.89} & \bf{0.90} & \bf{0.94} & 0.97 & 0.99 & {\bf 0.96} & {\bf 0.97} &563 & 550 & 455 & 740 & 733 & 631 & 811 & 516 & 1212 & 6211\\

& p\_gbridge & 1.00 & 1.00 & 1.00 & 1.00 & 1.00 & 0.99 & 1.00 & 1.00 & 1.00 &0 &0 &0 &0 &0 &0 &0 &0 &0 &0\\
\hline

\multirow{7}{*}{\rotatebox[origin=c]{90}{Month 6}} 
& WNET & 1.00 & 1.00 & 1.00 & \bf{0.94} & 0.98 & 1.00 & 1.01 & 0.99 & 1.00 &0 & 0 & 0 & 0 & 0 & 0 & 0 & 0 & 68 & 68\\

& WPCR & 1.35 & 1.09 & 1.22 & 1.50 & 1.42 & 1.43 & 1.42 & 1.51 & 1.48 &318 & 432 & 149 & 308 & 282 & 531 & 472 & 348 & 344 & 3184\\

& p\_lasso & 1.00 & 1.00 & 1.00 & 0.98 & 1.00 & 1.00 & 1.00 & 1.00 & 1.00 &0 & 0 & 0 & 0 & 0 & 0 & 0 & 0 & 0 & 0\\

& glasso & 0.98 & 1.00 & {\bf 0.97} & \bf{0.94} & 0.95 & 0.94 & 0.97 & 0.93 & \bf{0.95} &8 & 0 & 0 & 12 & 0 & 0 & 0 & 0 & 0 & 20\\

& gbridge & 1.00 & 1.00 & 1.00 & 1.00 & 1.00 & 1.00 & 1.00 & 1.00 & 1.00 &0 &0 &0 &0 &0 &0 &0 &0 &0 &0\\

& p\_glasso & \bf{0.89} & \bf{0.97} & {\bf 0.97} & \bf{0.94} & \bf{0.90} & \bf{0.92} & \bf{0.95} & \bf{0.91} & 1.00 &1257 & 1124 & 592 & 1169 & 1322 & 1049 & 1837 & 1279 & 2212 & 11841\\

& p\_gbridge & 1.00 & 1.00 & 1.00 & 1.00 & 1.00 & 0.99 & 1.00 & 1.00 & 1.00 &0 &0 &0 &0 &0 &486 &0 &0 &0 &486\\
\hline

\multirow{7}{*}{\rotatebox[origin=c]{90}{Month 12}} 
& WNET & 1.01 & 1.00 & 0.95 & 0.95 & 0.99 & 1.00 & 1.00 & 1.00 & 0.98 &0 & 0 & 0 & 3 & 0 & 0 & 0 & 0 & 81 & 84\\

& WPCR & 1.22 & 1.69 & 1.02 & 1.70 & 1.45 & 1.33 & 1.07 & 1.09 & 1.07 &1231 & 874 & 176 & 509 & 799 & 569 & 700 & 568 & 695 & 6121\\

& p\_lasso & 1.00 & 1.00 & 0.93 & 1.00 & 1.00 & 0.99 & 1.00 & 0.98 & 1.00 &0 & 0 & 0 & 0 & 0 & 0 & 0 & 0 & 11 & 11\\

& glasso & 1.00 & 1.00 & 0.97 & 0.94 & 1.00 & 0.99 & 0.99 & 0.99 & 0.95 &2 & 0 & 0 & 17 & 0 & 0 & 0 & 0 & 0 & 19\\

& gbridge & 1.00 & 1.00 & 1.00 & 1.00 & 1.00 & 1.00 & 1.00 & 1.00 & 1.00 &0 &0 &0 &0 &0 &0 &0 &0 &0 &0\\

& p\_glasso & \bf{0.96} & \bf{0.92} & \bf{0.89} & \bf{0.87} & \bf{0.91} & \bf{0.97} & {\bf 0.98} & \bf{0.90} & \bf{0.88} &1370 & 1329 & 976 & 1646 & 1718 & 702 & 2024 & 1915 & 2617 & 14297\\

& p\_gbridge & 1.00 & 1.00 & 1.00 & 1.00 & 1.00 & 0.99 & 1.00 & 1.00 & 1.00 &0 &0 &0 &0 &0 &896 &0 &0 &0 &896\\
\hline
\end{tabular}}
\caption{Left half of the Table shows the prediction MSE for ADNI data analysis, whereas the right half shows the number of significantly associated voxels, for each of the 9 axial slices. The bolded numbers imply significantly improved PMSE compared to other methods. }
\label{tab:adni_pmse}
\end{table}

\begin{figure}[hbt!]
\centering
\includegraphics[width=0.95\textwidth]{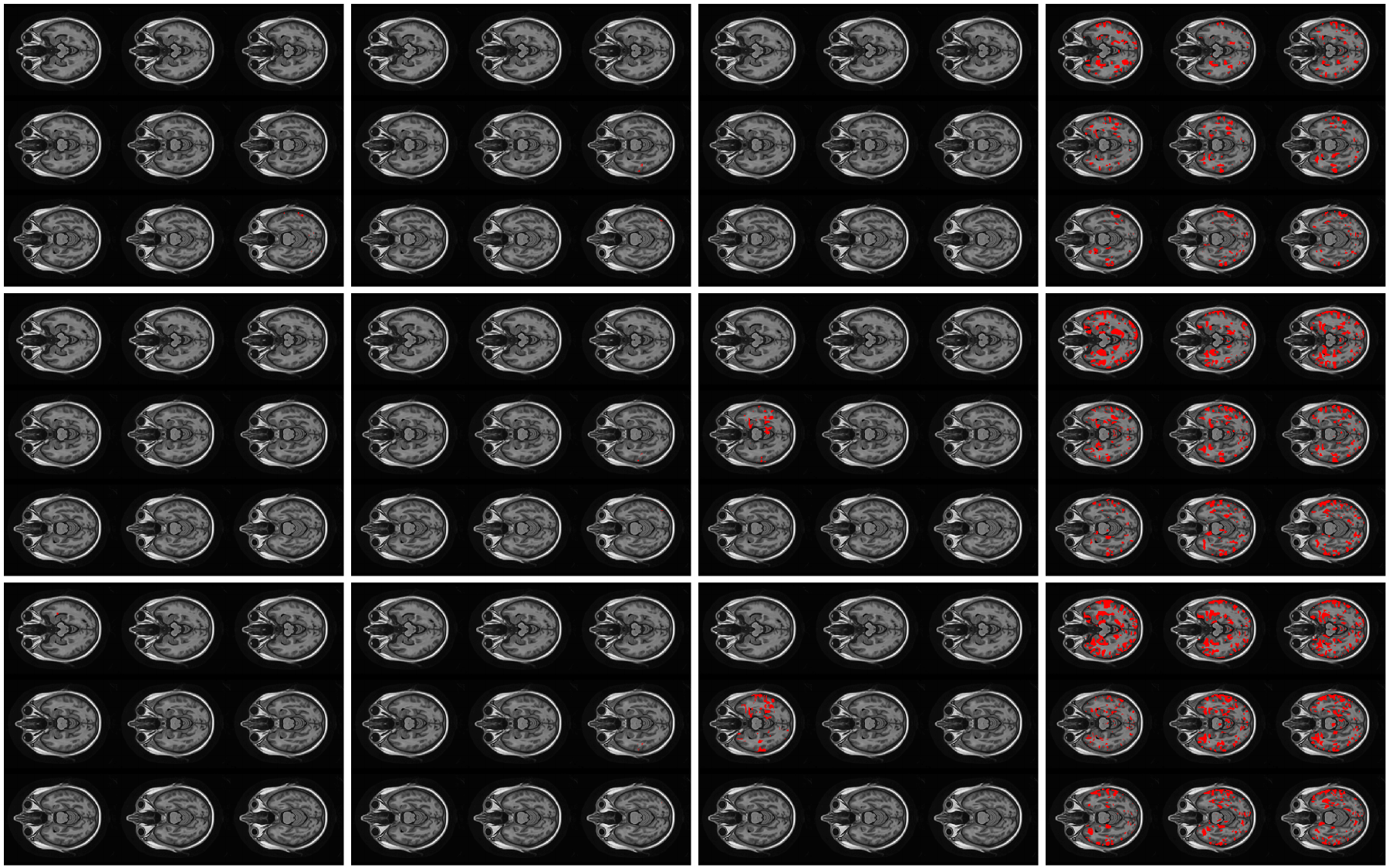}
\caption{Each sub-panel corresponds to association maps of the 9 axial slices. Columns 1-4 correspond to maps under the projected Lasso, group Lasso, projected group bridge and projected group Lasso methods respectively. The top, middle, and bottom rows correspond to maps at baseline, month 6 and month 12 respectively. The association maps for the remaining methods were reported in Figure 6 in Supplementary Materials.}
\label{fig:sigmaps}
\end{figure}

{\noindent \emph {Results}:} Table \ref{tab:adni_pmse} reports the out of sample prediction, and the association maps 
corresponding to the significant voxels are plotted in Figure \ref{fig:sigmaps}. 
Table \ref{tab:adni_pmse} also reports the number of significantly associated voxels across different methods. From the results, it is clear that while the projected group lasso with noise correction is able to detect significantly associated voxels in biologically interpretable regions (see below), all other competing methods report negligible or no significant associations after multiplicity corrections that is potentially due to the {\it attenuation to the null phenomenon} in the presence of noisy images. Our conjecture is that multi-task learning methods without noise correction lose their ability to detect common association patterns across longitudinal visits due to non-negligible noise-to-signal ratio in the brain images.
 The longitudinal association maps in Figure \ref{fig:sigmaps} illustrate the increase in the number of associated voxels over time under the projected group lasso (also see Table 2 in Supplementary Materials). It is seen that the significant voxels in early visits are highly likely to still be significant at later visits, and these are concentrated in the hippocampus, amygdala, and parahippocampal gyrus regions that is consistent with evidence in literature. 
The fusiform gyrus show significant associations that supports prior evidence linking this region to visual cognition deficits and work memory tasks in AD \citep{yetkin2006}.


It is also clear that all methods except the group lasso based approaches have inferior predictive performance, which highlight the advantage of pooling information across longitudinal images under group lasso. In addition, the projected group lasso approach has significantly improved prediction performance compared to the group lasso without noise correction for the vast majority of the 2-D slices. The prediction performance for the group bridge appears inferior compared to projected group lasso, which is potentially due to the lack of optimization bound guarantees under the projected gradient descent algorithm under the group bridge. Another potential explanation is that there is a large number of homogeneous signals across longitudinal visits, which is better addressed via the group lasso penalty compared to the group bridge penalty. We note that the prediction performance in terms of correlation between the observed and predicted test samples also illustrate strong gains under the projected group lasso (Table 5 in Supplementary Materials). Further, the projected group lasso method also has a superior prediction performance for the 3-D analysis (Table 6 in Supplementary Materials). Finally, we also fit a convolutional neural network (CNN) with standard architecture for the 2-D slices. However, with the limited samples available from the ADNI study and in the presence of noisy images, the CNN model demonstrated poor prediction power that was incomparable with our proposed methods (results not reported).

\section{Discussion}
\label{sec:discuss}
In this paper, we have proposed a novel approach for joint estimation of multiple scalar-on-image regression models involving noisy high-dimensional images. Although there is a rich literature on functional data analysis, the development for scalar-on-image regression methods is fairly recent, and existing methods in literature haven't addressed the question of mis-specification resulting from noisy images. Hence, the proposed methods are one of the first to address these issues via a novel M-estimation approach involving convex and non-convex group penalties that account for functional data mis-specifications. The implementation of the proposed methods are done via computationally efficient algorithms that are slightly slower than existing functional linear models that don't account for measurement error, but is still scalable to high-dimensional brain images.  The approach requires one to compute the noise covariance matrix that can be estimated from a validation dataset in ADNI analysis, and is largely robust to mis-specifications in the noise covariance. While we were able to establish optimization convergence results for convex grouped penalties, the corresponding results for non-convex grouped penalties are still an open problem with very limited prior literature \citep{fan2014strong}, and will be addressed in future work. 
The application of our proposed methods on the analysis of the ADNI T1-weighted MRI data provides a concrete example of the advantages of integrative learning via grouped penalties in multi-task learning over cross-sectional studies. Future work will include extending the proposed approach to other types of images, e.g. PET images and RAVENS maps.

\vskip 10pt

\spacingset{1.2}




{\noindent \bf \large Supplementary Materials}

Supplementary Materials contain proofs of theorems, additional results from simulations and ADNI analysis, and additional computational details.

\bibliography{literature}{}
\bibliographystyle{apalike}

\end{document}